\documentclass[seceq]{ptptex}

\usepackage{graphicx}
\usepackage{epsfig}
\usepackage{here}
\usepackage{wrapft}

\title{
The Origin of Space-Time as Seen from Matrix Model Simulations
}

\author{
Jun \textsc{Nishimura}%
}

\inst{
KEK Theory Center, High Energy Accelerator Research Organization,\\
1-1 Oho, Tsukuba, Ibaraki 305-0801, Japan
}

\abst{
The AdS/CFT correspondence, or more generally
the gauge/gravity duality, is a remarkable conjecture 
obtained from superstring theory with various D-brane backgrounds.
According to this conjecture, 
a higher-dimensional curved space-time emerges
from supersymmetric gauge theory in lower-dimensional flat space-time.
In the first part of this article, we review Monte Carlo studies
of U($N$) supersymmetric gauge theories, which confirmed
the gauge/gravity duality for various observables.
In particular, Monte Carlo results for
thermodynamic quantities enable us to understand
the microscopic origin of the black hole entropy 
associated with the dual geometry.
We also discuss results for Wilson loops and correlation functions,
which agree nicely with the predictions from the gravity side.
In the second part, we review recent developments in
a nonperturbative formulation of superstring theory,
which may be regarded as a counterpart of the lattice gauge theory
in QCD. In particular, we discuss Monte Carlo results for
the Lorentzian matrix model,
which suggest that (3+1)-dimensional expanding universe
emerges 
dynamically
from type IIB superstring theory in (9+1) dimensions
if one treats the theory nonperturbatively.
}

\newcommand {\SO}{\mathop{\rm SO}}
\newcommand {\SU}{\mathop{\rm SU}}
\newcommand {\beq} {\begin{equation}}
\newcommand {\eeq} {\end{equation}}
\newcommand {\beqa}{\begin{eqnarray}}
\newcommand {\eeqa}{\end{eqnarray}}
\newcommand {\n} {\nonumber}
\newcommand {\del} {\partial}
\newcommand {\tr}{{\rm tr\,}}

\newcommand {\Pf}{\mbox{Pf}}
\newcommand {\dd}{\mbox{d}}
\newcommand {\ee}{\mbox{e}}
\newcommand {\Imag}{\mbox{Im}}

\begin{document}

\maketitle

\section{Introduction}

Monte Carlo 
calculations based on lattice gauge theory
have been playing a central role
in studying various properties of QCD in a fully nonperturbative manner.
In this article, we would like to show that
similar developments are starting to take place
in superstring theory, which is a unified theory for
all the matters and the fundamental interactions including gravity.
As a theory of quantum gravity, the notion of space-time
had to undergo a revolutionary change.
It is by now widely appreciated that
matrices are the fundamental degrees of freedom
of superstring theory at the nonperturbative level,
and the space-time only emerges effectively
as a derived concept at low energy or at long distances.

One of the developments that manifest this idea
is the gauge/gravity duality. (See Ref.~\citen{Aharony:1999ti} 
for a comprehensive review.)
The original conjecture is known as the AdS/CFT correspondence,
which was put forward by Maldacena in 1997 \cite{AdS-CFT},
but it was soon generalized to non-conformal 
cases \cite{Itzhaki:1998dd}.
These dualities are 
arrived at
by considering two different
descriptions of D-brane backgrounds in superstring theory.
One is the gauge theory, which 
describes the open strings with both ends attached to the D-branes.
The other is a classical solution of supergravity, which describes 
the closed string degrees of freedom in the bulk 
sourced by the D-branes.
Such a remarkable statement that relates gauge theory and gravity theory is
possible precisely because superstring theory naturally
contains both of them.


When one considers the gauge theory at finite temperature,
the dual supergravity solution can have a geometry 
with an event horizon, which is characteristic to 
black holes (or ``black branes'', more generally).
It is well known that black holes have thermodynamic properties
although their microscopic origin has long been a mystery.
The gauge/gravity duality provides a very clear and explicit
answer to this problem.
The thermodynamic properties of a black hole
can be understood as those of the dual gauge theory,
which is considered to describe the
interior structure of the black hole.
%
Monte Carlo calculation
of thermodynamic quantities in the gauge theory
indeed 
reproduced the black hole 
thermodynamics \cite{AHNT,Hanada:2008ez,Catterall:2007fp}.
In particular, Ref.~\citen{Hanada:2008ez}
shows that the duality holds
including $\alpha '$ corrections, which represent
the effects of closed strings having finite length.
These results demonstrate that 
the gauge theory 
describes correctly the quantum space-time structure
at the center of the black hole.

The correspondence between gauge theory and gravity 
was extended to the level of operators,
which is of particular importance in using the duality
to study strongly coupled gauge theories by
much simpler calculations in supergravity.
In particular,
explicit prescriptions for obtaining Wilson loops \cite{maldloop} 
and correlation functions \cite{Gubser:1998bc} were proposed.
The predictions obtained by such prescriptions have also been 
confirmed by direct Monte Carlo calculations on the gauge theory 
side \cite{Hanada:2008gy,HNSY1,HNSY2}.

Monte Carlo studies mentioned above
deal with the simplest case
of D0-branes, which corresponds to one-dimensional 
U($N$) super Yang-Mills theory (SYM)
with 16 supercharges.
In fact, it is possible to extend these works to 
higher dimensions by extending the idea of the large-$N$ 
reduction \cite{Eguchi:1982nm}
to a curved space \cite{Ishii:2008ib},
while circumventing a well-known problem \cite{Bhanot:1982sh} 
in the original idea.
This unconventional regularization scheme,
as opposed to the lattice,
enables calculations respecting supersymmetry maximally.
Some preliminary 
results are obtained \cite{Nishimura:2009xm,Honda:2010nx,Honda:2011qk}
for the case of D3-branes, which corresponds to four-dimensional 
${\cal N}=4$ 
U($N$) SYM.



An important aspect of the gauge/gravity duality is that
higher-dimensional curved space-time emerges
from the gauge theory in lower-dimensional flat space-time.
The extra spatial dimensions are actually described in gauge theory
by the scalar fields in the adjoint representation 
of U($N$) gauge group,
which are represented by $N\times N$ matrices.
This is an example
of ``emergent space'',
which appears in various contexts of string theory \cite{Seiberg:2006wf}.
One of the important open questions is whether one can extend the idea
to emergent space-time instead of just space.
Here we quote a sentence from Seiberg's 
lecture \cite{Seiberg:2006wf} 
in 2005: 
\emph{Understanding how time emerges will undoubtedly shed new light
on some of the most important questions in
theoretical physics including the origin of the Universe.}



The second development we would like to review in this article
concerns 
a 
nonperturbative 
formulation of superstring theory,
which is considered as a counterpart of the lattice gauge theory
in the case of QCD.
In particular, we discuss recent Monte Carlo results for a Lorentzian
matrix model \cite{Kim:2011cr}, which 
actually show that \emph{the emergent space-time} seems to be naturally realized.
%
Back in 1996, 
Ishibashi, Kawai, Kitazawa and Tsuchiya
proposed
a matrix model,
which is called the type IIB matrix model,
as a nonperturbative definition of type IIB superstring theory
in ten dimensions \cite{IKKT}.
The model can be obtained formally from 
SYM
that appears in the aforementioned
examples of the gauge/gravity duality by
dimensionally reducing them to zero dimension.
Thus one obtains
ten bosonic matrices and sixteen fermionic matrices, 
which do not have dependence on space-time coordinates.
The entire space-time is expected 
to emerge dynamically from the ten bosonic matrices.
%


Until quite recently, however, 
it was common to study the type IIB matrix model
after making a ``Wick rotation''. 
%
This amounts to replacing the Hermitian matrix $A_0$
in the temporal direction
by $A_0 = i A_{10}$,
and treating the Hermitian matrix $A_{10}$ 
on equal footing as the matrices $A_i$ ($i=1,\cdots 9$)
in the spatial directions.
The Euclidean model obtained in this way has 
manifest SO(10) symmetry, and it
is well defined as Monte Carlo studies
with small matrices demonstrate \cite{Krauth:1998xh}.
In fact the partition function was proven to be finite
for arbitrary matrix size \cite{AW}.
In Ref.~\citen{AIKKT},
perturbative expansion around
the diagonal configurations $A_\mu = {\rm diag}(x_{1\mu} , \cdots ,
x_{N\mu} )$ 
was studied and
the low-energy effective
theory for the diagonal elements
was discussed.
In particular, it was speculated that configurations with the $N$ points
$\{ \vec{x}_{i} ; i=1,\cdots , N\}$ distributed on a four-dimensional
hypersurface in ten-dimensional Euclidean space
may be favored due to some nontrivial interactions
in the low-energy effective theory.
If that really happens, it implies that 
the SO(10) symmetry is spontaneously
broken down to SO(4) and 
that four-dimensional space-time is
generated dynamically.

Monte Carlo studies of the Euclidean model is difficult
due to the sign problem since the Pfaffian that appears from
integrating out the fermionic matrices is complex in general.
Monte Carlo studies of the model omitting the phase of the Pfaffian
show that the SO(10) symmetry is not spontaneously broken \cite{Ambjorn:2000dx}.
In fact 
the phase of the Pfaffian 
has an effect of favoring lower-dimensional configurations \cite{NV}.
It is expected that such an effect can be studied by Monte Carlo 
simulation in the near future 
by using a new method to overcome 
the sign problem \cite{Anagnostopoulos:2001yb,Anagnostopoulos:2010ux}

As an alternative approach to this issue,
the Gaussian expansion method was proposed \cite{Nishimura:2001sx}.
Recently,
the free energy was calculated 
by assuming that 
the SO($d$) symmetry ($2 \le d \le 7$) remains unbroken,
and it was found that $d=3$ gives 
the minimum \cite{Nishimura:2011xy}.
Another important observation from the Gaussian expansion method
was that
the extent of space-time in the extended $d$ directions
and that in the shrunken $(10-d)$ directions 
turn out to have a finite ratio
even in the large-$N$ limit \cite{Nishimura:2011xy}.
While these results reveal interesting dynamical 
properties of the Euclidean
model, which can also be understood intuitively
from the viewpoint of 
the low-energy effective theory, 
the connection to our real space-time is not very clear.

Motivated by these results for the Euclidean model,
Kim, J.N.\ and Tsuchiya \cite{Kim:2011cr} 
studied
the type IIB matrix model \emph{without making the Wick rotation}.
The action has an SO(9,1) symmetry instead of SO(10).
The reason why no one dared to study this \emph{Lorentzian model} 
before beyond the classical level \cite{Steinacker:2011wb}
was that the bosonic action $S_{\rm b}$ is not positive definite
unlike the Euclidean case, and therefore
the system seemed to be highly unstable.
Moreover,
the bosonic action
becomes a pure phase in the integrand of the partition function as
in the path integral formulation of 
quantum field theories in Minkowski space.
Therefore it seemed just impossible to get anything sensible
out of the Lorentzian model without making a Wick rotation.

On the other hand, it is known from many 
examples that the Wick rotation is subtle
in theories including gravity.
For instance, the Lorentzian quantum gravity
has been pursued within 
the dynamical triangulation approach \cite{Ambjorn:2005qt}
motivated from earlier studies of the 
Euclidean gravity, and the results
turned out to be quite different.
More recently, the worm hole scenario as a solution
to the cosmological constant problem was
reconsidered in the Lorentzian quantum gravity \cite{Kawai:2011rj},
and the results provided a consistent picture,
which was not available in the original Euclidean formulation.

The crucial trick to make the Lorentzian matrix model
accessible by Monte Carlo simulation is to
integrate out the scale factor
of the bosonic matrices first,\footnote{The same
procedure was also used in Ref.~\citen{Krauth:1998xh} 
for simulating the Euclidean model.}
which essentially converts
the phase factor $e^{i S_{\rm b}}$ into a constraint
$S_{\rm b} \approx 0$.
This is possible since the action of the type IIB matrix model
is homogeneous\footnote{Note that the gauge theories 
before dimensional reduction
do not have this property due to the existence of 
the derivative terms.}
with respect to the matrices.
The model one obtains in this way 
does not have 
the sign problem
since the Pfaffian that appears from integrating out fermionic
matrices is real in the Lorentzian case.

First of all, 
Monte Carlo studies confirmed that
the Lorentzian matrix model is not well defined as it is.
It was found that 
the extents in the temporal and spatial directions,
which are represented by 
$\frac{1}{N}\tr (A_0)^2$ and $\frac{1}{N}\tr (A_i)^2$, 
respectively, both tend to diverge.
One therefore has to put cutoffs on these quantities.
However, it turned out that the two cutoffs can be
removed in the large-$N$ limit in such a way that the
results scale in $N$. 
The theory thus obtained
turned out to have no parameters other than the scale parameter, which
can be naturally identified as the string scale.
This is a highly nontrivial property of 
the Lorentzian matrix model, which 
supports its validity
as a nonperturbative formulation of superstring theory.


Another important observation in the Monte Carlo studies is that
the eigenvalue distribution of the matrix $A_0$ representing the 
time direction extends 
as one takes the large-$N$ 
limit explained above.
Supersymmetry plays a crucial role here.
(For the bosonic model, the eigenvalue distribution of the matrix 
$A_0$ has finite extent 
even in the large-$N$ limit.)

Moreover, by making an SU($N$) transformation 
in such a way that 
the temporal matrix $A_0$ is diagonalized,
the nine-dimensional space represented by $A_i$ ($i=1,\cdots ,9$)
is found to exhibit a sensible ``time
evolution''.
%
In fact, the space remains small and SO(9) symmetric
from the infinite past until some ``critical time'', at which only three directions
start to expand rapidly. 
This implies that
the rotational SO(9) symmetry in the spatial directions 
is spontaneously broken down to SO(3) at the critical time,
which may be identified as ``the birth of our Universe''.
The so-called initial condition problem is not an issue 
in the present framework since even 
the time evolution is an emergent concept. 
The mechanism of the spontaneous symmetry breaking (SSB)
relies crucially on the noncommutativity of space,
and it seems totally different from the Euclidean case, 
in which the SSB is considered
to be caused by the phase of the Pfaffian.
In the Lorentzian model, the Pfaffian is real as we mentioned above.



The rest of this article is organized as follows.
In section \ref{sec:MCgg}
we review the developments related to the gauge/gravity duality.
In particular, we discuss its direct tests based on Monte Carlo studies
of supersymmetric gauge theories.
In section \ref{sec:nonpert-form}
we review the developments related to a nonperturbative formulation
of superstring theory. In particular, we discuss
how (3+1)-dimensional expanding universe emerges from
the Lorentzian matrix model.
In section \ref{sec:sum-all} we conclude 
with a summary and future prospects.

\section{Monte Carlo studies of the gauge/gravity duality}
\label{sec:MCgg}




The gauge/gravity duality \cite{Aharony:1999ti} is conjectured by considering
D-brane backgrounds in superstring theory.
D-branes are known to be consistent backgrounds in superstring theory,
and they may be considered as a counterpart of solitons in field
theory. 
Note that solitons do not appear in perturbative expansion around the
trivial vacuum, and they are considered as nonperturbative objects.
From that point of view,
D-branes are expected to capture some nonperturbative 
aspects of string theory.
Indeed D-branes played a crucial role
in finding the duality web of superstring/M theory and in 
constructing a nonperturbative formulation of superstring/M theory.

\begin{wrapfigure}{r}{6.6cm}
\centerline{\includegraphics[width=6.6cm]{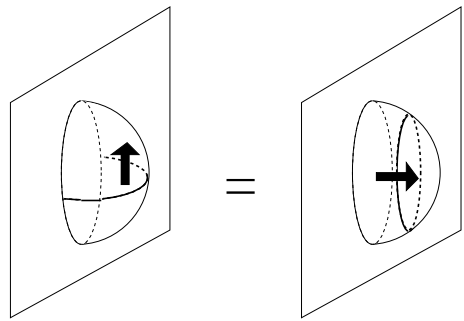}}
\caption{
On the left, an open string attached to the D-brane
is propagating along it.
On the right, the same process is
viewed as
emission of a closed string from the D-brane.
}
\label{dbrane}
\end{wrapfigure}
D-brane can extend in $p$ dimensions, and it is characterized
as a hypersurface on which strings can end on.
(``D'' stands for the Dirichlet boundary condition
imposed at the boundary of the worldsheet of an open string.)
Let us consider an open string attached to the D-brane
propagating along it.
In Fig.~\ref{dbrane} on the left, 
we describe such a process diagrammatically.
If one slices the diagram in the orthogonal direction,
one notices that the same process can be viewed as emission of
a closed string.
This is an example of the so-called open-string/closed-string duality.
Note here that an open string and a closed string
include a gauge particle and a graviton, respectively, as massless modes.
This observation lies at the heart of the gauge/gravity duality.

In order to formulate the gauge/gravity duality,
we need to consider $N$ D-branes lying on top of each other
and take a particular low-energy limit so that the open strings attached
to the D-branes and the closed strings in the bulk are decoupled.
Then one has two independent descriptions of D-branes.
One is the $(p+1)$-dimensional U($N$) SYM,
which describes the open strings attached to the D-branes.
The other is a solution to supergravity,
which describes the closed string degrees of freedom in the bulk
sourced by the D-branes.
Note that the bulk ten-dimensional space-time is curved 
since D-branes emit gravitons.
In order for the supergravity description to be valid,
one needs to take certain limits on the gauge theory side.
\begin{itemize}
\item The so-called 
't Hooft large-$N$ limit with fixed $\lambda\equiv g_{\rm YM}^2 N$.
(The string loop corrections are suppressed by $1/N$.)
\item The large-$\lambda$ limit.
(The $\alpha '$ corrections, which are 
due to strings having finite extent,
are suppressed by some powers of $1/\lambda$.)
\end{itemize}
A stronger version of the conjecture claims that
the gauge/gravity duality holds including the string loop corrections 
and the $\alpha '$ corrections. If that is the case, one can say that
the gauge theory, which is well defined for all $N$ and $\lambda$,
actually defines superstring theory on the particular
curved background nonperturbatively.


In addition to this perspective, there are various reasons why
the gauge/gravity duality are considered very interesting.
First of all, it is a realization of an old idea 
by 't Hooft \cite{'tHooft:1973jz},
which states that the large-$N$ gauge theory is equivalent to 
some classical string theory, although in those days people may not
have anticipated that the string theory actually lives in a curved 
space-time.
It is interesting that the curved space-time
emerges from a gauge theory in a flat space.
This aspect of the duality is often referred to as 
the \emph{emergent space} \cite{Seiberg:2006wf}.
In the gauge/gravity duality, one typically obtains 
the anti-de Sitter space. 
If one considers the gauge theory at finite temperature,
one can have a black-hole-like geometry \cite{Itzhaki:1998dd,Witten:1998zw}.
Therefore, one may explain the microscopic origin of the 
black hole thermodynamics in terms of gauge theory.
One can also use the duality in the opposite direction,
and study strongly coupled gauge theories,
which are relevant to hadron and condensed matter physics,
from a curved space-time.

Since the gauge/gravity duality is a strong-weak duality,
it is important to study gauge theories in the strongly coupled regime. 
Monte Carlo simulation can be a powerful tool for such purposes.
However, the problem is that the gauge theories we are interested in
have supersymmetry, which is broken by the lattice.
This can be seen immediately if one recalls the supersymmetry algebra
$\{Q , \bar{Q} \} \propto P_\mu $, where the generators for translation
appear on the right hand side. Since the translational symmetry is broken
by the lattice regularization, one necessarily breaks supersymmetry.
The best one can do is to restore supersymmetry in the continuum limit
by fine-tuning some parameters in the action, which requires a lot of
efforts, however.

Recently there are considerable developments 
in ``lattice supersymmetry'',
which can be categorized into two classes.
One is the construction of lattice actions with various symmetries.
For instance, one can preserve one supercharge by using the so-called
topological twist.
The other one, which we discuss in what follows, is
to use a regularization different from the lattice.

In the case of D0-branes, which corresponds to the 
supersymmetric gauge theory in 1 dimension 
with 16 supercharges,
one can regularize the theory using momentum cutoff
after fixing the gauge appropriately \cite{Hanada-Nishimura-Takeuchi}.
Black hole dynamics have been reproduced including the $\alpha'$
corrections \cite{AHNT,Hanada:2008ez},
and the gauge/gravity duality for the Wilson loops
\cite{Hanada:2008gy}
and the correlation functions \cite{HNSY1,HNSY2} 
has been confirmed.

Notably, one can extend this approach 
to 3d and 4d gauge theories
by using the idea of large-$N$ reduction \cite{Ishii:2008ib}.
In the 4d case, the gauge theory becomes superconformal and the 
number of supersymmetries enhances from 16 to 32.
This superconformal theory is interesting on its own right,
but it is also studied intensively in the context of 
the AdS/CFT correspondence, which is a typical case of the
gauge/gravity duality \cite{AdS-CFT}.
The non-lattice simulation of the 4d superconformal theory
requires no fine-tuning,
unlike the previous proposals based on the 
lattice regularization \cite{latticeSUSY_N4}.
As we will see, preliminary results for the Wilson loops
and the correlation functions are 
promising \cite{Nishimura:2009xm,Honda:2010nx,Honda:2011qk}.

\subsection{Non-lattice simulation of 1d SYM with 16 supercharges}
\label{sec:non-lattice}


Let us start with the D0-brane case, which corresponds
to the 1d U($N$) SYM with 16 supercharges.
The action is given by $S = S_{\rm b} + S_{\rm f}$, where
\beqa
S_{\rm b} &=&  \frac{1}{g^2} \ \int_{0} ^{\beta}  dt \
\tr \left\{  \frac{1}{2} \Bigl(D X_i(t) \Bigr)^2
- \frac{1}{4} [X_i(t) , X_j (t)] ^2
\right\}   \ , \\
S_{\rm f}&=&  \frac{1}{g^2} \ \int_{0} ^{\beta}  dt \
\tr \left\{  \frac{1}{2}\Psi_\alpha D \Psi _\alpha
- \frac{1}{2} \Psi_\alpha (\gamma_i)_{\alpha\beta}
 [X_i , \Psi_\beta]  \right\} \ .
\label{bfss-action}
\eeqa
The covariant derivative is denoted as
$D = \partial_t  - i \,  [A (t), \ \cdot \ ]$.
$X_j(t) \ (j=1, \cdots , 9)$
and $\Psi_\alpha(t) \ (\alpha=1, \cdots , 16) $
are $N\times N$ Hermitian matrices, and the theory has SO(9) symmetry.
When we are interested in finite temperature,
we impose periodic boundary conditions on $X_j(t)$ 
and anti-periodic boundary conditions on $\Psi_\alpha(t)$.
Then the temperature is given by $T\equiv \beta^{-1}$,
where $\beta$ is the extent in the Euclidean time ($t$) direction.
The 't Hooft coupling constant is defined by
$\lambda \equiv g^2 N$, which has the dimension of mass cubed.
The physics of the system is determined only by the dimensionless 
coupling constant $\lambda_{\rm eff} \equiv \frac{\lambda}{T^3}$.
Therefore one can take $\lambda=1$ without loss of generality.
With this convention, the low $T$ regime corresponds to the 
strongly coupled regime, which is expected to have the dual gravity
description \cite{Itzhaki:1998dd}, 
whereas the high $T$ regime is essentially weakly coupled,
and the high temperature expansion (HTE) is applicable \cite{HTE}.

In non-lattice simulation \cite{Hanada-Nishimura-Takeuchi},
we introduce
an upper bound on the Fourier mode as
$X_i  (t) = \sum_{n=-\Lambda}^{\Lambda} 
\tilde{X}_{i ,n} \ee^{i \omega n t}$, where
$\omega = \frac{2\pi}{\beta}$,
and similarly for the fermions.
This idea does not work usually because it breaks gauge invariance.
(Recall that the Fourier mode is not a gauge invariant concept.)
However, in 1d, one can fix the gauge nonperturbatively in the 
following way.
We first take the static diagonal gauge
$A(t) = 
\frac{1}{\beta}
{\rm diag}(\alpha_1 , \cdots ,
\alpha_N )$, in which the gauge field
is constant in time and diagonal.
By following the usual Faddeev-Popov procedure, one obtains 
\beq
S_{\rm FP}
= - \sum_{a<b}
2 \ln  \left|\sin \frac{\alpha_a -\alpha_b}{2} \right|
\eeq
as a term to be added to the action.
The above gauge choice does not fix the gauge symmetry completely,
and there is a residual symmetry given by
\beq
\alpha_a \mapsto  \alpha_a + 2 \pi \nu_a \ , 
\quad \quad
\tilde{X}_{i, n}^{ab} \mapsto  \tilde{X}_{i , n-\nu_a + \nu_b}^{ab} \ , 
\quad \quad 
\tilde{\Psi}_{\alpha, n}^{ab} 
\mapsto  \tilde{\Psi}_{\alpha , n-\nu_a + \nu_b}^{ab} \ ,
\eeq
which represents a topologically nontrivial gauge transformation
corresponding to the gauge function
$g(t) = {\rm diag} (\ee ^{i \omega \nu_1 t } , \cdots ,
\ee ^{i \omega \nu_N t } )$.
This residual gauge symmetry can be fixed by imposing
$-\pi < \alpha_a \le \pi$.
One can then introduce the Fourier mode cutoff $\Lambda$.
Since there is no UV divergence in this 1d model,
one can take the $\Lambda \rightarrow \infty$ limit naively,
and one retrieves the original
gauge theory with 16 supercharges.

The system with finite $\Lambda$
can be simulated efficiently \footnote{Strictly speaking,
the Pfaffian one obtains from integrating out the fermions
is complex in general. However, there is numerical evidence \cite{HNSY2}
that the phase can be omitted without altering the results.
Complete justification is left for future investigations.}
by using the standard RHMC algorithm \cite{Clark:2003na}.
In particular, the Fourier acceleration \cite{Catterall:2001jg}
can be implemented without
extra cost since we are dealing with the Fourier modes directly as
the fundamental degrees of freedom. This is crucial in reducing the
critical slowing down 
at large $\Lambda$.
(The same theory is also studied using the standard lattice
approach \cite{Catterall:2007fp}.
However, from the results obtained so far,
the non-lattice simulations seem to be far more
efficient in obtaining the continuum limit.)

\begin{figure}[htb]
\begin{center}
\begin{tabular}{cc}
\epsfig{file=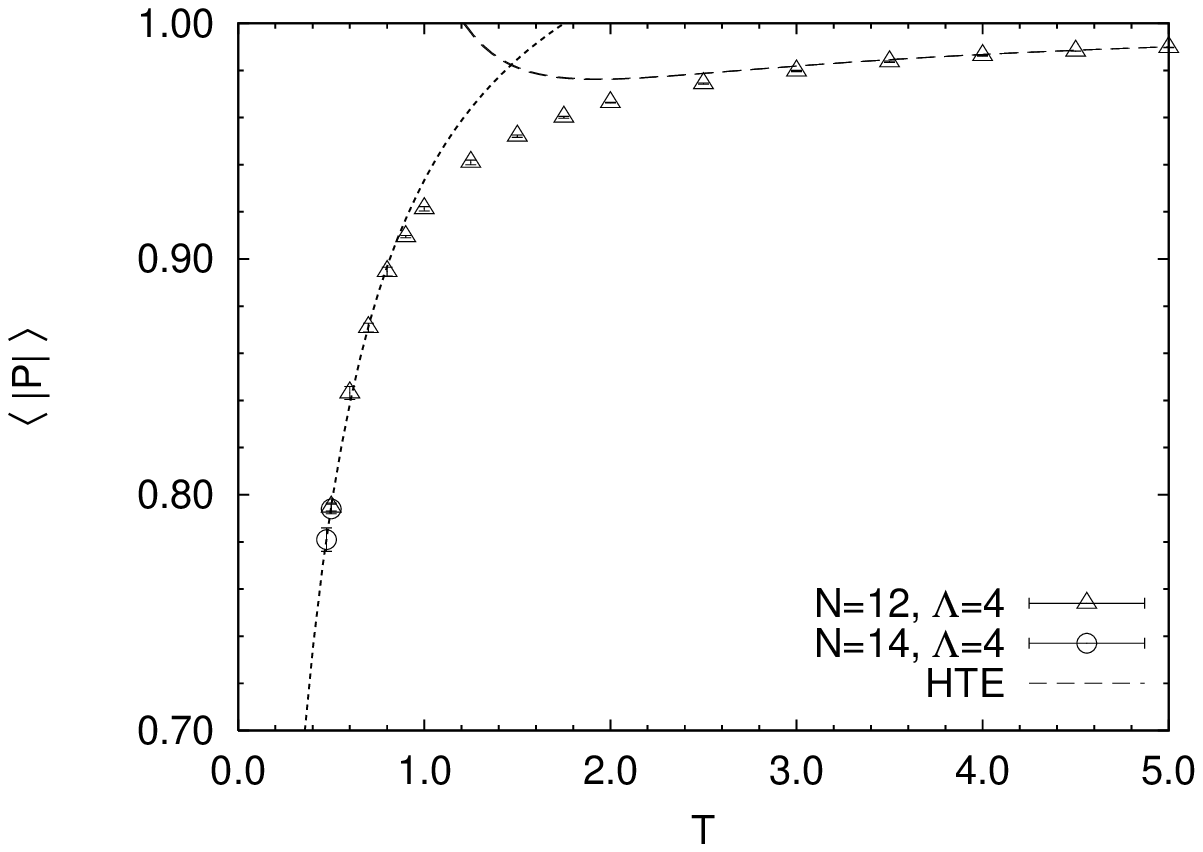,width=.47\textwidth} &
\epsfig{file=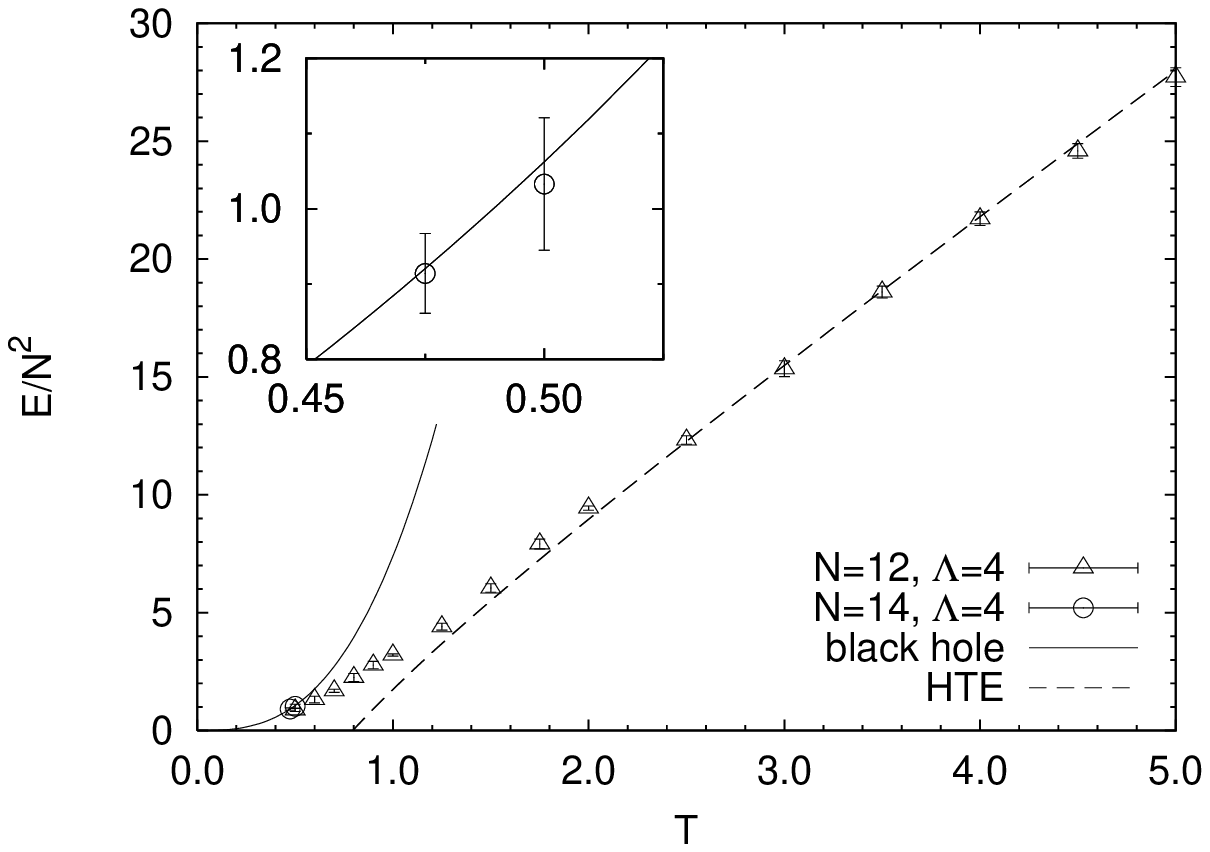,width=.47\textwidth} \\
\end{tabular}
\end{center}
\caption{%
(Left) 
The Polyakov line
is plotted against 
$T$.
The dashed line represents the result
of HTE
up to the next leading order for $N=12$ \cite{HTE}.
The dotted line represents a fit to
eq.\ (\protect\ref{deconf-P})
with $a=0.15$ and $b=0.072$.
(Right) 
The energy
(normalized by $N^2$)
is plotted against $T$.
The dashed line represents the result
obtained by HTE
up to the next leading order 
for $N=12$ \cite{HTE}.
The solid line represents
the asymptotic
power-law behavior at small $T$
predicted by the gauge/gravity duality.
}
\label{fig:old}
\end{figure}

Let us first discuss the phase structure that appears
when one changes the temperature.
As is well known, the Polyakov line serves as an order parameter
for the spontaneous breaking of the center symmetry.
Fig.~\ref{fig:old} (Left) shows the results \cite{AHNT}.
At high temperature the data agree nicely with 
the HTE \cite{HTE} 
including the next-leading order.
As the temperature decreases below $T \sim 3$, the data start to
deviate, and at low temperature below $T \sim 0.9$, the data can be
fitted to the characteristic behavior of the ``deconfined phase''
\beq
\langle |P| \rangle = \exp \left(-\frac{a}{T}+b \right) \ .
\label{deconf-P}
\eeq
In the temperature regime investigated, we find no phase transition.
This is in sharp contrast to the bosonic model
\cite{latticeBFSS,Aharony:2004ig,Kawahara:2007fn},
which undergoes
a phase transition to the ``confined phase''
at $T \sim 0.9$.
The absence of the phase transition is consistent
with analyses on the gravity 
side \cite{Witten:1998zw,Barbon:1998cr}.


\subsection{Black hole thermodynamics from 1d SYM}

Let us turn to a quantitative prediction from the gauge/gravity
duality.
Given the dual geometry, one can use Hawking's theory 
of black hole thermodynamics to obtain various thermodynamic
relations such as \cite{Klebanov:1996un}
\beq
\frac{1}{N^2}\left( \frac{E}{\lambda^{1/3}}
\right)= c
 \, \left( \frac{T}{\lambda^{1/3}} \right)
^{14/5}  \ , \quad \quad
c =  \frac{9}{14}
\left\{ 4^{13} 15^{2} \left ( 
\frac{\pi}{7}
\right) ^{14}
 \right\} ^{1/5} = 7.41 \cdots \ .
\label{gravity-leading}
\eeq
The gauge/gravity duality
predicts that this should be reproduced
by 1d SYM in the large-$N$ limit at low $T$ \cite{Itzhaki:1998dd}.
The importance of this prediction is that,
if it is true,
it explains the microscopic origin of the black hole thermodynamics,
meaning that the 1d SYM provides the quantum description of 
the states inside the black hole. 

In Fig.~\ref{fig:old} (Right)
we plot the internal energy \cite{AHNT}, which is defined by
$E = \frac{\partial}{\partial \beta} (\beta {\cal F})$
in terms of the free energy $\mathcal{F}$.
At $T\gtrsim 3$
the data agree with 
the HTE \cite{HTE}. 
As one goes to lower temperature,
the data points approach the solid line, which 
represents the result (\ref{gravity-leading})
obtained from the 10d black hole.
(See Refs.~\citen{KLL} for earlier studies based on
the Gaussian approximation.)

\begin{figure}[tb]
\begin{center}
\begin{tabular}{cc}
\epsfig{file=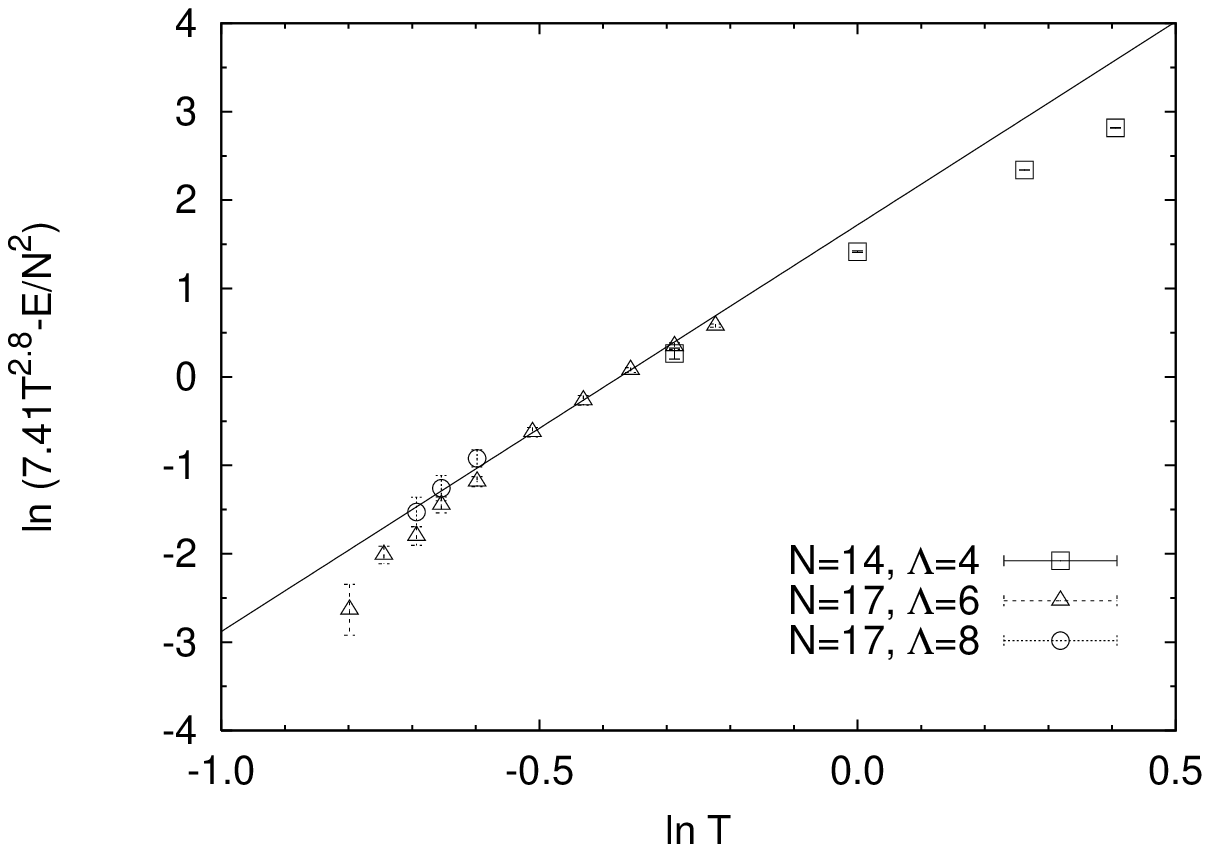,width=.47\textwidth} &
\epsfig{file=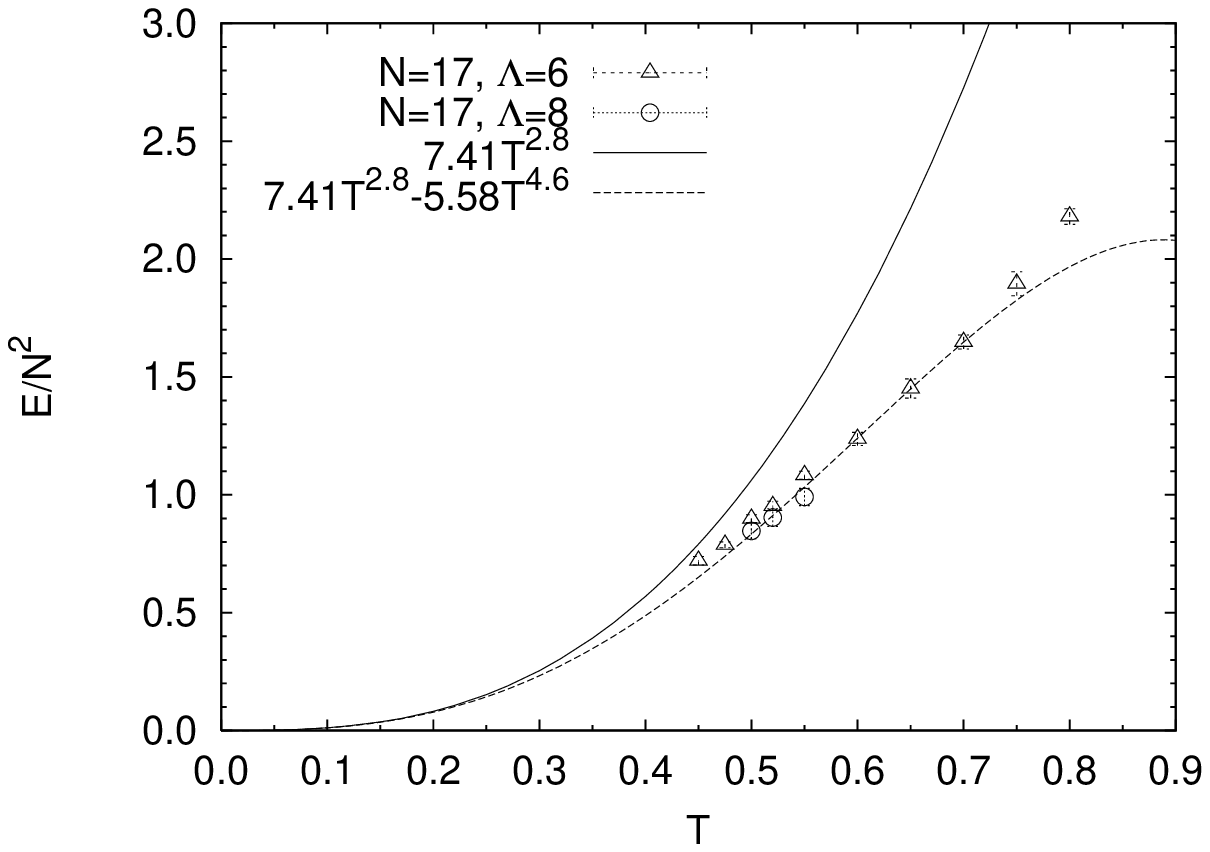,width=.47\textwidth} \\
\end{tabular}
\end{center}
\caption{%
(Left) 
The deviation of the internal energy
$\frac{1}{N^2} E$ from the leading term
$7.41 \, T^{\frac{14}{5}}$
is plotted against the temperature
in the log-log scale for $\lambda=1$.
The solid line represents 
a fit to a straight line with the slope 4.6
predicted from the $\alpha '$ corrections 
on the gravity side.
(Right) 
The internal energy $\frac{1}{N^2} E$
is plotted against $T$ for $\lambda=1$.
The solid line represents
the leading asymptotic behavior at small $T$
predicted by the gauge/gravity duality.
The dashed line represents
a fit to the behavior (\ref{sub-leading})
including the subleading term
with $C=5.58$.
}
\label{fig:energy}
\end{figure}


From Fig.~\ref{fig:old} (Right) alone,
it is not clear whether the gauge theory
results continue to follow the line predicted from gravity at lower $T$.
In fact, simulations at lower $T$ are difficult, since one has to
increase $\Lambda$ proportionally to $1/T$, and at the same time
one has to increase $N$ to avoid the run-away behavior 
due to finite $N$ \cite{AHNT}.
Instead of lowering $T$, 
we determined
the power of the subleading term as \cite{Hanada:2008ez}
\beq
\frac{1}{N^2}\left( \frac{E}{\lambda^{1/3}}
\right)= c
 \, \left( \frac{T}{\lambda^{1/3}} \right)
^{14/5}  
- C \,  \left( \frac{T}{\lambda^{1/3}} \right)
^{23/5} 
\ ,
\label{sub-leading}
\eeq
from gravity.
This was derived by considering
higher derivative corrections
in the supergravity action due to the effects
of strings having finite extent ($\alpha ' $ corrections).
The coefficient $C$ of the subleading term
is calculable in principle, but it requires 
the full information of the higher derivative corrections, which are
yet to be determined.
By using (\ref{sub-leading}), however,
we can already make a nontrivial test of 
the gauge/gravity duality \cite{Hanada:2008ez}.
In Fig.~\ref{fig:energy} (Left) we plot the discrepancy
$7.41 T^{14/5}-E/N^2$ against $T$ in the log-log scale,
which reveals that the power of the subleading term is indeed 
consistent with the predicted value $23/5=4.6$.
In Fig.~\ref{fig:energy} (Right) we find that the data at $T\lesssim 0.7$ 
can be nicely fitted to the form (\ref{sub-leading}) with $C=5.58$.
Note also that the $\Lambda=6$ data seem to suffer 
from some finite $\Lambda$ effects at low $T$. 
From this point of view, we consider that 
the $\Lambda=4$ data points at low $T$
in Fig.~\ref{fig:old} (Right),
which seem to be on the curve of the leading order 
result from gravity, also suffer from finite $\Lambda$ effects.
Now we know that actually the subleading term 
in (\ref{sub-leading})
should be taken into account 
for precise agreement.

The fact that the internal energy goes to zero as 
a power of $T$ towards $T=0$ is closely related to 
the existence of the threshold bound state in the 1d gauge
theory \cite{Smilga:2008bt}.
The leading power-law behavior $T^{14/5}$ can actually be understood 
by considering that excitations around the threshold bound state
have energy of the order of $N^{-5/9}$ as suggested from the effective
Hamiltonian.
In the case of 1d SYM with four and eight supercharges,
it is considered that the threshold bound state does not exist.
Indeed, Monte Carlo studies show that
the internal energy decreases faster as $\sim e^{-c \, T}$ as 
$T$ goes to zero \cite{Hanada:2010jr}.

\subsection{Schwarzschild radius from Wilson loop in 1d SYM}

As another prediction from the gauge/gravity duality, 
let us consider the Wilson loop, which winds around the temporal 
direction once.
Unlike the usual Polyakov line, 
we consider the one involving the adjoint scalar as
\beq
W \equiv \frac{1}{N} \,
\tr \,  {\cal P} \exp \left[ i \int_0^\beta
 dt \{ A(t) + i \, n_i \, X_i(t) \} \right]
\ ,
\label{Maldloop-def}
\eeq 

\begin{wrapfigure}[18]{r}{6.6cm}
\centerline{\includegraphics[width=6.6cm]{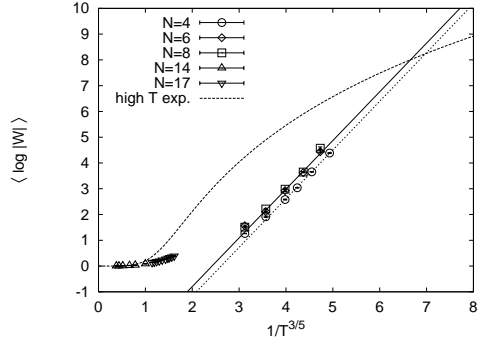}}
\caption{
The plot of
$\langle\log|W|\rangle$ for $\lambda=1$
against $T^{-3/5}$. 
The cutoff $\Lambda$ is chosen as follows: 
$\Lambda=12$ for $N=4$; $\Lambda=0.6/T$ for $N=6,8$; 
$\Lambda=4$ for $N=14$; $\Lambda=6$ for $N=17$.
The dashed line represents the results
of the 
HTE
up to the next-leading order
for $N=14$, which are obtained by applying the method 
in Ref.~\citen{HTE}.
}
\label{fig:logW}
\end{wrapfigure}
\noindent where $n_i$ is a unit vector in 9d, which can be chosen arbitrarily
due to the SO(9) invariance.
This object can be calculated on the gravity side by 
considering the minimal surface spanning the loop
in the dual geometry \cite{maldloop}.
For the present model, the result is given by \cite{Hanada:2008gy}
\beq
\ln W = 
\frac{\beta R_{\rm Sch}}{2 \pi \alpha ' }
 = \kappa
 \left(\frac{T}{\lambda^{1/3}} \right)^{-3/5} \ ,
\label{Wilson-loop}
\eeq
where $R_{\rm Sch}$ is the Schwarzschild radius of the dual black hole
geometry and
\beq
\kappa = \frac{1}{2 \pi}\left\{ 
\frac{16 \sqrt{15} \pi^{7/2}}{7} 
\right\}^{2/5}  =  1.89\cdots \ .
\label{kappa-predicted}
\eeq
In Fig.~\ref{fig:logW} 
we plot the log of the Wilson loop \cite{Hanada:2008gy} against
$T^{-3/5}$ anticipating (\ref{Wilson-loop}).
Indeed, at low temperature (to the right on the figure),
we find that the data points can be fitted nicely to a straight line
with a slope 1.89 in precise agreement with (\ref{kappa-predicted}).
The solid line corresponds to
$\langle \log |W| \rangle = 1.89 \, T^{-3/5} - 4.58 $,
where the existence of the constant term 
can be understood as $\alpha ' $ corrections.
This result demonstrates that one can extract the information of the dual 
geometry such as the Schwarzschild radius from the gauge invariant
observable (\ref{Maldloop-def}).


\subsection{Correlation functions in 1d SYM}

One can also predict correlation functions from the gravity side.
This was done more than ten years ago by Sekino and Yoneya \cite{Sekino:1999av}
applying the Gubser-Klebanov-Polykov-Witten prescription
\cite{Gubser:1998bc}
to the present 1d SYM case.

\begin{wrapfigure}{r}{6.6cm}
\centerline{\includegraphics[width=5cm,angle=270]{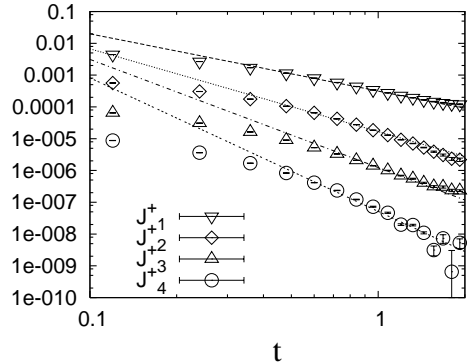}}
\caption{
The log-log plot of the correlator 
$\Bigl \langle J^+_{\ell}(t)\, J^+_{\ell}(0) \Bigr\rangle$
with  $\ell=1,2,3,4$ for $N=3$.
The cutoff parameters are chosen as $\beta=4$ and $\Lambda=16$.
The straight lines represent the power-law behavior predicted
by the gauge/gravity correspondence.
}
\label{fig:Jp}
\end{wrapfigure}
For instance, let us consider an operator
\beq
\mathcal{O}_{\ell} = 
\frac{1}{N} \,
{\rm Str} 
\Bigl(
F_{ij}X_{i_1}\cdots X_{i_\ell}
\Bigr) \ , 
\label{Jplus}
\eeq
where $F_{ij}\equiv -i \, [X_i,X_j]$ and 
${\rm Str}$ represents the symmetrized trace treating $F_{ij}$ 
as a single unit.
The two-point correlation function of this operator
is predicted as
\beq
\langle \mathcal{O}_{\ell}(t) 
\mathcal{O}_{\ell}(0) \rangle
\sim \frac{1}{|t|^{p}} \ , 
\quad
p = \frac{4\ell}{5} +1
\label{SYpredict}
\eeq
at $\lambda ^{-1/3} \ll |t| \ll
\lambda ^{-1/3} N^{10/21}$.
In Fig.~\ref{fig:Jp} 
we plot the two-point correlation function \cite{HNSY2}
for $\ell = 1,2,3,4$, which
agrees precisely with the predicted power-law behavior.\footnote{As 
a closely related work, Refs.~\citen{Hiller:2000nf}
presented a numerical analysis based
on the discrete light-cone quantization
for the $(1+1)$-dimensional case in the large-$N$ limit.
The two-point correlation function of the stress-energy tensor
has been calculated, and the expected power-law behavior has been confirmed.}
What is rather surprising is that the agreement
is observed even for such small $N$ as $N=3$.
In particular, the power-law behavior
seems to extend to the far infrared regime, in which the
supergravity calculations become invalid.
This fact may have important implications on M theory interpretation
of the same model.
See Ref.~\citen{HNSY2} for more details
as well as results for other operators.

In general, the correlation functions of operators 
which correspond to supergravity modes
on the gravity side show power-law behavior as predicted by
the gauge/gravity duality.
Some operators are predicted to show unusual infrared
diverging behavior such as $|p|^{-6/5}$ in the momentum space.
The gauge/gravity duality is confirmed by Monte Carlo calculations 
even 
in such cases.
The operators corresponding to stringy excited modes 
are also studied on the gravity side \cite{ASY}, 
and they are predicted
to have correlation functions with an exotic behavior
$e^{-c \, t^{3/5}}$. Monte Carlo results for these operators
are indeed consistent with such a behavior.


\subsection{Extension to higher dimensions 
based on the large-$N$ reduction}
\label{sec:higherD}

In this section we discuss how one can extend the works
in the previous sections to higher dimensions.
Respecting supersymmetry becomes more nontrivial 
in higher dimensions, 
and we use the idea of the large-$N$ reduction, which relates
gauge theories in higher dimensions to those in lower dimensions
in the 't Hooft large-$N$ limit.
This, in particular, enables us to perform Monte Carlo studies of
the D3-brane case, which corresponds
to the four-dimensional ${\cal N}=4$ SYM.

Here we use a novel large-$N$ reduction, which was proposed in
Ref.~\citen{Ishii:2008ib}
to study ${\cal N}=4$ SYM on $R\times S^3$.
The reduced model in this case is obtained by
collapsing the $S^3$ to a point,
and it is given by the action
\begin{eqnarray}
S_{\rm PW}
&=& \frac{1}{g_{\rm PW}^2}
\int
dt \, \tr 
\left[\frac{1}{2}(D_tX_M)^2-\frac{1}{4}[X_M,X_N]^2
+\frac{1}{2}\Psi D_t \Psi
-\frac{1}{2}\Psi\gamma_M[X_M,\Psi] \right.\nonumber\\
&~& \quad \quad \left.+\frac{\mu^2}{2}(X_i)^2
+\frac{\mu^2}{8}(X_a)^2 +i\mu\epsilon_{ijk}X_iX_jX_k
+i\frac{3\mu}{8}\Psi \gamma_{123}\Psi \right] \ ,
\label{pp-action}
\end{eqnarray}
where the parameter $\mu$ is 
related to the radius of $S^3$ as $R_{S^3}=\frac{2}{\mu}$,
and the covariant derivative is defined by
$D_t=\partial_t-i[A, \ \cdot \ ]$,
where $A(t)$ as well as $X_M(t)$ and $\Psi (t)$ 
is an $N\times N$ Hermitian matrix.
The range of indices is given by 
$1 \le M,N \le 9$, $1 \le i,j,k \le 3$ and $4 \le a \le 9$, respectively.
The model has the SU$(2|4)$ symmetry with 16 supercharges.

The one-dimensional supersymmetric gauge theory (\ref{pp-action}) 
is called the plane wave matrix model (PWMM) 
%
or the BMN matrix model \cite{Berenstein:2002jq}\footnote{Properties 
of this model at finite temperature are studied at weak coupling
\cite{Kawahara:2006hs,eff_PWMM}
and at strong coupling \cite{Catterall:2010gf}.}.
It is nothing but the 1d SYM
discussed in the previous sections,
plus some mass deformation, which 
preserves 16 supersymmetries of the undeformed theory.

The PWMM possesses many discrete vacua 
representing multi fuzzy spheres, which are given explicitly by
\begin{equation}
X_i=\mu \bigoplus_{I=1}^{\nu}
\Bigl( L_i^{(n_I)}\otimes {\bf 1}_{k_I} \Bigr) \ 
\quad \mbox{with}\ \ \sum_{I=1}^{\nu}n_Ik_I=N \ ,
 \label{background}
\end{equation}
where $L_i^{(r)}$ are
the $r$-dimensional irreducible representation of the SU$(2)$ algebra 
$[L_i^{(r)},L_j^{(r)}]=i \, \epsilon_{ijk} \, L_k^{(r)}$.
These vacua preserve the SU$(2|4)$ symmetry, and are all degenerate.

In order to retrieve the planar ${\cal N}=4$ SYM on $R \times S^3$,
one has to pick up a particular background from (\ref{background}),
and consider the theory (\ref{pp-action}) around it.
Let us consider the vacuum defined by
\begin{equation}
k_I=k \ , \quad n_I=n+I-\frac{\nu+1}{2} \quad \quad
\mbox{for\ \ $I =1, \cdots , \nu$} \ ,
\label{our background}
\end{equation}
and take the large-$N$ limit in such a way that
\begin{eqnarray}
k\rightarrow \infty \ , \
\frac{n}{\nu} \rightarrow \infty \ , \ 
\nu\rightarrow\infty \ ,
\quad \mbox{with} \;\;
\lambda_{\rm PW} \equiv \frac{g_{\rm PW}^2 k}{n}
\; \; \mbox{fixed} \ .
\label{limit}
\end{eqnarray}
%
Then the resulting theory is claimed
\cite{Ishii:2008ib} to be equivalent\footnote{
See Refs.~\citen{earlier_novel} for earlier studies that led to 
this proposal. 
This equivalence was 
checked at finite temperature in the weak coupling regime \cite{eff_PWMM}.
It has also been extended to general group manifolds and 
coset spaces \cite{group-coset}.}
to the planar limit of ${\cal N}=4$ SYM on $R \times S^3$
with the 't Hooft coupling constant given by
\beq
\lambda_{\rm SYM} =2\pi^2 \lambda_{\rm PW}(R_{S^3})^3 
=\frac{16 \pi^2 k }{n}
\frac{g_{\rm PW}^2}{\mu^3} \ .
\label{lambda-sym}
\eeq
%
In practice, we use (\ref{background})
with (\ref{our background}) as the initial configuration
and check that no transition to other vacua occurs 
during the simulation.

The above equivalence may be viewed as
an extension of the large-$N$ reduction \cite{Eguchi:1982nm},
which asserts that the large-$N$ gauge theories can be 
studied by dimensionally reduced models.
It is known that
the original idea for theories compactified on a torus
can fail due to
the instability of the U(1)$^D$ symmetric vacuum
of the reduced model \cite{Bhanot:1982sh}.
This problem is avoided in the novel proposal
since the PWMM is a massive theory and the vacuum preserves the maximal SUSY.
This regularization respects 16 supersymmetries, which is half of 
the full superconformal symmetry of ${\cal N}=4$ SYM on $R\times S^3$.
Since any kind of UV regularization breaks the conformal symmetry,
this regularization is optimal from the viewpoint of preserving SUSY.

Since the parameter $g_{\rm PW}^{2}$ in the action (\ref{pp-action})
can be scaled out by appropriate redefinition of fields and parameters,
we take $g_{\rm PW}^{2}N=1$ without loss of generality
as 
in Refs.\ \citen{AHNT,Hanada:2008gy,Hanada:2008ez,HNSY1,HNSY2}.
In this convention one finds from eq.\ (\ref{lambda-sym}) that
the small (large) 
$\mu$ region in the PWMM corresponds to the strong (weak) coupling 
region in the 4d $\mathcal{N}=4$ SYM.

\subsection{Wilson loops in 4d $\mathcal{N}=4$ SYM}
\label{sec:wloop}

Let us consider the following type of Wilson loop
\begin{equation}
W(C) = \frac{1}{N}\tr \mathcal{P} 
   \exp{\oint_C ds \left( iA_\mu^{R^4} 
\dot{x}^\mu(s) +\left| \dot{x}^\mu (s) \right| X_a^{R^4} 
\theta_a  \right) }  \ ,
\end{equation}
where $\dot{x}^\mu (s)\equiv \frac{dx^\mu (s)}{ds}$ and $\theta_a$ 
is a constant which satisfies $\theta_a \theta_a =1$.
The fields $A_\mu^{R^4}$ and $X_a^{R^4}$ represent 
the gauge field and the six scalars, respectively,
in 4d $\mathcal{N}=4$ SYM on $R^4$. 
Due to the particular way in which the scalars appear,
one can obtain predictions from the gravity side based on 
the AdS/CFT correspondence as \cite{maldloop}
\begin{equation}
\lim_{N\rightarrow\infty ,\lambda_{\rm SYM}\rightarrow\infty} 
\Big\langle W(C) \Big\rangle_{\rm SYM} 
=  e^{-S(C)} \ ,
\end{equation}
where $S(C)$ represents the area of the minimal surface spanning
the loop $C$ on the boundary of the AdS space.

For the circular Wilson loop $W(C_{\rm circ})$, 
which is a (1/2-)BPS operator,
there is an exact result on the gauge theory side,
which is obtained by summing up planar ladder diagrams \cite{circular}
or by using the localization method \cite{Pestun:2007rz}.
The result is given by
\begin{eqnarray}
\lim_{N\rightarrow\infty} 
\Big\langle W(C_{\rm circ}) \Big\rangle_{\rm SYM} 
&=& \sqrt{\frac{2}{\lambda_{\rm SYM}}} \, 
I_1 \Big(\sqrt{2\lambda_{\rm SYM}} \Big) 
\label{localization} \\
&\simeq & \frac{e^{\sqrt{2\lambda_{\rm SYM}}}}
{\left( \frac{\pi}{2}\right)^{1/2}(2\lambda_{\rm SYM})^{3/4}} 
          \quad \quad  \rm{for}\ \lambda_{\rm SYM}\gg 1  \ ,
\label{strong}
\end{eqnarray}
where $I_1 (x)$ is the modified Bessel function of the first kind.
The result is independent of the radius of the circle,
which is a consequence of the scale invariance of $\mathcal{N}=4$ SYM. 
At strong coupling it agrees with the result obtained 
from the dual geometry \cite{gravC}
$S(C_{\rm circ})=-\sqrt{2\lambda_{\rm SYM}}$.
This is an explicit example of the AdS/CFT correspondence.
We use the exact result (\ref{localization}) for arbitrary $\lambda_{\rm SYM}$
to test our calculation method.


The Wilson loop in $\mathcal{N}=4$ SYM 
can be calculated in PWMM in the following way.
When we perform the conformal mapping from $R^4$ to $R\times S^3$,
the radial and angular directions are mapped to 
the time and $S^3$-directions, respectively.
Therefore, an arbitrary loop on a plane in $R^4$ is mapped to a loop
on $R\times S^3$, which can be projected to a great circle on $S^3$.
Such a Wilson loop can be represented in the large-$N$ reduced model as
\begin{equation}
W_{\rm red}(C) = \frac{1}{N}\tr \mathcal{P} 
   \exp{\oint_C ds \left( i A_0 \frac{dt}{ds}  +i X_i e_\mu^i  \dot{x}^\mu (s) 
        +\left| \dot{x}_\mu (s) \right| X_a \theta_a  \right) }  \ ,
\label{op_red}
\end{equation}
where 
$e_j^i (x^\mu (s))$ 
is the dreibein on $S^3$.

\begin{wrapfigure}{r}{6.6cm}
\centerline{\includegraphics[width=6cm]{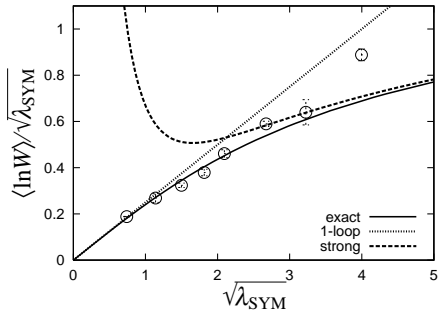}}
  \caption{The log of the circular Wilson loop normalized 
by $\sqrt{\lambda_{\rm SYM}}$ is plotted against $\sqrt{\lambda_{\rm SYM}}$.
The solid line represents the exact result (\protect\ref{localization}). 
The dashed line represents the behavior (\protect\ref{strong})
at strong coupling, whereas the dotted line represents 
the leading perturbative behavior 
$\ln{\langle W\rangle}\simeq \frac{1}{4}\lambda_{\rm SYM}$.
}
\label{fig:circle}
\end{wrapfigure}
The expectation value of this operator is related 
to the average of the original Wilson loop 
as \cite{Ishii:2007sy}
\begin{equation}
\Big\langle W(C) \Big\rangle_{\rm SYM}
= \Big\langle W_{\rm red}(C) \Big\rangle \ ,
\label{relation-op}
\end{equation}
where $\langle \cdots \rangle$ on the right-hand side denotes the expectation value 
in the large-$N$ reduced model (PWMM).
In the case of the circular Wilson loop, the relation (\ref{relation-op})
was confirmed by reproducing the SYM result (\ref{localization})
from the reduced model to all orders in perturbation theory
assuming that non-ladder diagrams do not contribute \cite{Ishiki:2011ct}.


In Fig.~\ref{fig:circle} we present our preliminary results 
for the circular Wilson loop \cite{Honda:2011qk}.
We have performed the $\Lambda\rightarrow\infty$ extrapolation
using $\Lambda$ = 6,8,10,12
assuming that finite $\Lambda$ effects are O($1/\Lambda$). 
The extent in the time direction is 
fixed to $\beta =5$. 
The parameters describing the background (\ref{our background})
are chosen to be $(n ,\nu )=(\frac{3}{2},2 )$, and
we performed an extrapolation to $k=\infty$ using the data 
for $k = 2,3,4,5$ assuming that the finite-k effects are
O($1/k^2$).
We also plot the exact result (\ref{localization}).
Except for the data point at $\sqrt{\lambda_{\rm SYM}}=4$,
the agreement with the exact result is promising.
Note, in particular, that we already
start to observe a bent from the weak coupling behavior 
towards the strong coupling behavior.
This is remarkable considering the rather small matrix size.
We consider this as a result of the fact that our formulation
respects sixteen supersymmetries.



\subsection{Correlation functions in 4d $\mathcal{N}=4$ SYM}
\label{sec:corr-fun-sym}

In this section
we consider chiral primary operators (CPOs) 
as simple examples of $1/2$ BPS operators
in 4d $\mathcal{N}=4$ SYM,
and present Monte Carlo results for 
their correlation functions \cite{Honda:2010nx}.
In particular, we find that the two-point and three-point functions
agree with the free theory results up to overall constant factors
even at fairly strong coupling.
Moreover the ratio of the overall factors agrees with the prediction
of the AdS/CFT correspondence.\footnote{There are 
also Monte Carlo studies of the 4d ${\cal N}=4$ SYM based on 
matrix quantum mechanics of 6 bosonic commuting matrices
\cite{Berenstein_sim}, which give results consistent with
the AdS/CFT for the three-point functions of CPOs.}

\begin{figure}[htbp]
\begin{center}
\begin{tabular}{cc}
\epsfig{file=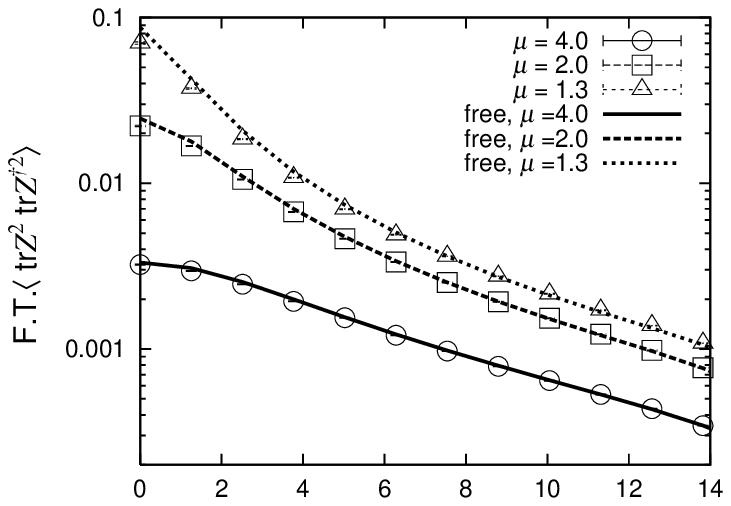,width=.47\textwidth} &
\epsfig{file=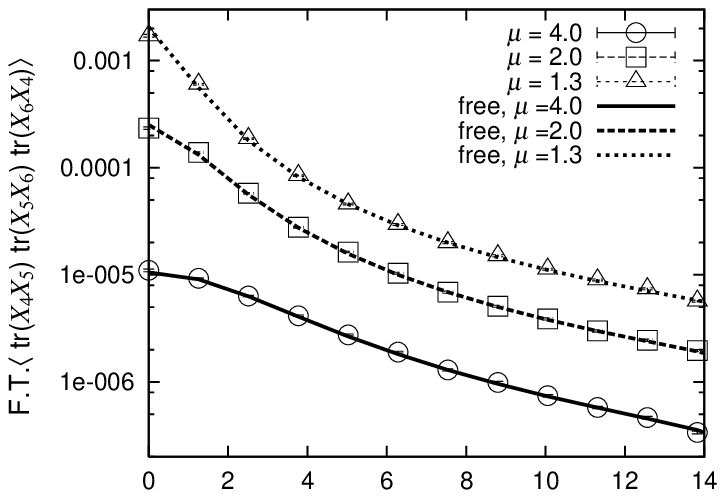,width=.47\textwidth} \\
\end{tabular}
\end{center}
  \caption{(Left) The two-point function 
$\Big\langle \tr\widetilde{Z^{2}}(p)\ 
\tr \widetilde{Z^{\dag 2}}(-p) \Big\rangle $ 
is plotted 
in the log scale.
The curves represent 
the corresponding free theory results
multiplied by 0.919, 0.799, 0.647 for $\mu=4.0,2.0,1.3$,
respectively.
(Right) The three-point function
$\Big\langle \tr\left( 
\widetilde{{X}_{4} X_{5}}(p)\right) \, 
\tr\left( \widetilde{ X_{5}X_{6} }(0)\right) \, 
\tr\left( \widetilde{X_{6}X_{4}}(-p)\right) \Big\rangle $
is plotted 
in the log scale. 
The curves represent 
the corresponding free theory results
multiplied by 0.850, 0.716, 0.491 for $\mu=4.0,2.0,1.3$,
respectively.
}
  \label{fig:ch2}
\end{figure}

Let us consider the CPOs given by
\begin{equation}
\mathcal{O}^{R^{4}}_{\Delta}(x)= 
T_{a_{1}\cdots a_{\Delta}}\ 
\tr\left( X^{R^{4}}_{a_{1}}X^{R^{4}}_{a_{2}}\cdots X^{R^{4}}_{a_{\Delta}}(x) 
\right) \ ,
\end{equation}
where $T_{a_{1}\cdots a_{\Delta}}$ is a symmetric traceless tensor
and $X_{a}^{R^{4}}$ represents 
the six scalars in 4d $\mathcal{N}=4$ SYM on $R^{4}$.
Thanks to the conformal symmetry, the forms of two-point and three-point 
functions of the CPOs are determined as
\begin{eqnarray}
\left\langle\mathcal{O}^{R^{4}}_{\Delta}(x_{1}) 
\mathcal{O}^{R^{4}}_{\Delta}(x_{2}) \right\rangle 
&=& c_{\Delta} \left\langle\mathcal{O}^{R^{4}}_{\Delta}(x_{1}) 
\mathcal{O}^{R^{4}}_{\Delta}(x_{2}) \right\rangle _{\rm free} \ , 
\nonumber \\
\left\langle\mathcal{O}^{R^{4}}_{\Delta_{1}}(x_{1}) 
\mathcal{O}^{R^{4}}_{\Delta_{2}}(x_{2}) 
\mathcal{O}^{R^{4}}_{\Delta_{3}}(x_{3}) \right\rangle 
&=& c_{\Delta_{1}\Delta_{2}\Delta_{3}} \left\langle
\mathcal{O}^{R^{4}}_{\Delta_{1}}(x_{1}) 
\mathcal{O}^{R^{4}}_{\Delta_{2}}(x_{2}) 
\mathcal{O}^{R^{4}}_{\Delta_{3}}(x_{3}) \right\rangle _{\rm free} \ , 
\label{O-Ofree}
\end{eqnarray}
where $c_{\Delta}$ and $c_{\Delta_{1}\Delta_{2}\Delta_{3}}$ are 
over-all constants depending on $\lambda_{\rm SYM}$ in general,
and $\langle\cdots \rangle _{\rm free}$ denotes the results of free theory.
The analysis on the gravity side suggests \cite{Lee:1998bxa}
\begin{equation}
\left. \frac{c_{\Delta_{1}\Delta_{2}\Delta_{3}}}
{\sqrt{c_{\Delta_{1}}c_{\Delta_{2}}c_{\Delta_{3}}}} 
\right|_{N\rightarrow\infty,\lambda_{\rm SYM}\rightarrow\infty}
=\left. \frac{c_{\Delta_{1}\Delta_{2}\Delta_{3}}}
{\sqrt{c_{\Delta_{1}}c_{\Delta_{2}}c_{\Delta_{3}}}} 
\right|_{N\rightarrow\infty,\lambda_{\rm SYM}\rightarrow 0}=1
\quad \mbox{for}\ \ ^{\forall}\Delta_{i} \ . 
\label{ratio_gravity}
\end{equation}

In order to relate the above operators
to those in the PWMM, 
we first perform the conformal mapping\footnote{ 
The metrics of $R^4$ and $R\times S^3$
are related as
$ds^2_{R^4}  = dr^2+r^2d\Omega_3^2  =e^{\mu t}ds^2_{R\times S^3}$, 
where $r=\frac{2}{\mu}e^{\frac{\mu}{2}t}$.
The transformation of the CPOs is given by
$\mathcal{O}^{R\times S^{3}}_{\Delta}=
e^{\frac{\Delta}{2}\mu t} \mathcal{O}^{R^{4}}_{\Delta} $.}
from $R^{4}$ to $R\times S^{3}$.
Then the $M$-point functions of the 
CPO $\mathcal{O}^{R\times S^{3}}_{\Delta_{i}}$ on $R\times S^{3}$ 
are related to those 
in PWMM as
\begin{eqnarray}
&~& \int\frac{d\Omega_{3}^{(1)}}{2\pi^{2}}\cdots 
\int\frac{d\Omega_{3}^{(M)}}{2\pi^{2}}
\left\langle \mathcal{O}_{\Delta_{1}}^{R\times S^{3}}
( t_{1},\Omega_{3}^{(1)}) \cdots 
\mathcal{O}_{\Delta_{M}}^{R\times S^{3}}
( t_{M},\Omega_{3}^{(M)} ) \right\rangle   \\
&=& \frac{1}{n^M\nu} \left\langle \mathcal{O}_{\Delta_1}^{\rm PW}(t_1)
\cdots \mathcal{O}_{\Delta_M}^{\rm PW}(t_M) \right\rangle  \ ,
\label{correspondence}
\end{eqnarray}
where we have defined \cite{Ishii:2008ib}
$\mathcal{O}^{\rm PW}_{\Delta}(t)= T_{a_{1}\cdots a_{\Delta}}\ 
\tr\Big( X_{a_{1}}X_{a_{2}}\cdots X_{a_{\Delta}}(t) \Big)$. 

We calculate the two-point functions
$\Big\langle\tr Z^{2}(t_{1}) \, \tr Z^{\dag 2}(t_{2})\Big\rangle$,
where 
$Z=\frac{1}{\sqrt{2}}(X_{4}+iX_{5})$,
and the three-point functions
$\Big\langle \tr
\Big( X_{4}X_{5}(t_{1})\Big) \, \tr
\Big( X_{5}X_{6}(t_{2})\Big) \, \tr \Big( X_{6}X_{4}(t_{3}) \Big) 
\Big\rangle$.
The CPOs we consider here have $\Delta =2$, and
%
the AdS/CFT predicts
$c_{222}=c_{2}^{3/2}$,
which we test by Monte Carlo calculations.

\begin{wrapfigure}{r}{6.6cm}
\centerline{\includegraphics[width=6cm]{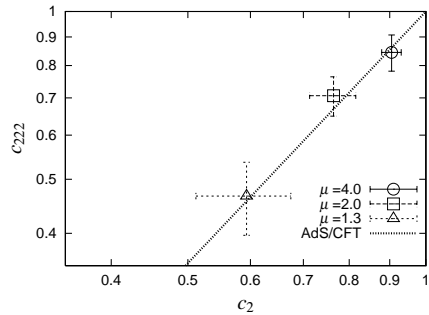}}
  \caption{The overall constants corresponding to 
$c_{2}$ and $c_{222}$ 
in eq.\ (\protect\ref{O-Ofree})
are plotted in the log-log scale.
The straight line presents the relation
$c_{222}=c_{2}^{3/2}$ predicted by the AdS/CFT.
}
  \label{fig:ch23}
\end{wrapfigure}


The parameters describing the background (\ref{our background})
are chosen as $n=\frac{3}{2},\nu =2,k=2$, which corresponds 
to the matrix size $N=6$.
The values of $\mu$ we use are
$\mu=4.0,2.0,1.3$, which correspond 
to $\lambda_{\rm SYM}\simeq 0.55,4.39,16.0$, respectively,
in the chosen background.
Thus we cover a wide range of the coupling constant.
The regularization parameters in the $t$-direction are
taken as $\beta =5.0$ and $\Lambda =12$ for all cases.

In Fig.\ \ref{fig:ch2} (Left)
we plot the two-point function\footnote{The Fourier transform
of an operator $\mathcal{O}(t)$ is defined
as $\tilde{\mathcal{O}}(p)=\frac{1}{\beta} \int_{0}^{\beta}dt \, 
\mathcal{O}(t) \, e^{-ipt}$.}
$\Big\langle \tr \widetilde{Z^{2}}(p) \, 
\tr \widetilde{Z^{\dag 2}}(-p) \Big\rangle$.
We find that the results agree well --- up to overall constants 
depending on $\mu$ ---
with
the corresponding free theory results,
which are obtained analytically
by just switching off the interaction terms
in the reduced model with the same regularization parameters.
In Fig.~\ref{fig:ch2} (Right) we show similar results 
for the three-point function defined by 
$\Big\langle \tr\left( 
\widetilde{{X}_{4} X_{5}}(p)\right) \, 
\tr\left( \widetilde{ X_{5}X_{6} }(0)\right) \, 
\tr\left( \widetilde{X_{6}X_{4}}(-p)\right) \Big\rangle $.


We can extract the the overall constants corresponding to
$c_{2}$ and $c_{222}$ in eq.\ (\ref{O-Ofree}) from
Fig.~\ref{fig:ch2}.
In Fig.\ \ref{fig:ch23}
we plot the overall constants 
obtained in this way for three values of $\mu$.
The data points represent the mean value of 
the upper and lower bounds.
We find that our results 
for various coupling constants
lie on the straight line which represents
the prediction $c_{222}=c_{2}^{3/2}$
from the AdS/CFT.
Our results therefore suggest that the relation
(\ref{ratio_gravity}) holds also at intermediate 
coupling constants.

\section{Nonperturbative formulation of superstring theory}
\label{sec:nonpert-form}

Superstring theory not only provides a most natural candidate for a
consistent theory of quantum gravity but also enables unified
description of all the interactions and the matters.
A crucial problem is that
we do not yet have a well-established nonperturbative formulation,
which would be needed in addressing dynamical issues
such as the determination of space-time dimensionality.\footnote{See 
Ref.~\citen{Brandenberger:1988aj} for a well-known scenario 
based on string-gas cosmology.}

%
%
%


In the 1990s, there was a remarkable progress in understanding the
nonperturbative aspects of superstring theory based on
D-branes.
Most importantly, it was noticed that large-$N$ matrices
are the appropriate
microscopic degrees of freedom which are useful in formulating
superstring theory in a nonperturbative
manner \cite{BFSS,IKKT,Dijkgraaf:1997vv}.
In particular, the type IIB matrix model
was proposed as a nonperturbative formulation of
type IIB superstring theory
in ten-dimensional space-time \cite{IKKT}.
%
It was also realized that the five types of
superstring theory in ten dimensions
are just different descriptions of the same theory.
Therefore,
it was speculated
that the type IIB matrix model
actually describes the unique underlying theory, although
it takes the form
that has
explicit connection
to perturbative type IIB superstring theory \cite{IKKT,Fukuma:1997en}.

In the type IIB matrix model, the space-time is represented 
\emph{dynamically}
by the eigenvalue distribution
of ten bosonic $N\times N$
traceless Hermitian matrices \cite{AIKKT}.
%
We first discuss the Euclidean model, which is obtained
by making a Wick rotation.
Then we 
discuss a recent work on the Lorentzian model,
which shows that (3+1)-dimensional expanding universe emerges
dynamically \cite{Kim:2011cr}.

\subsection{Definition of the Euclidean model}

%
%

 
%
%
%

The type IIB matrix model can be obtained formally
by the zero-volume limit of 
$D=10$ 
$\SU(N)$ pure super Yang-Mills theory.
The partition function of the Euclidean model is given by 
\begin{eqnarray}
  Z &=& \int \dd A \, \dd \Psi
\, e^{-S_{\rm b} - S_{\rm f}} \ , 
\label{eq:10dpf} \\
  S_{\rm b} &=& - \frac{1}{4 g^2}  \tr [ A_\mu, A_\nu ]^2  \ , 
\label{eq:sb} \\
  S_{\rm f} &=& - \frac{1}{2 g^2} 
\tr\left(\Psi_\alpha ({\cal C} \Gamma^\mu)_{\alpha\beta}
    [ A_\mu, \Psi_\beta ] \right) \ . 
\label{eq:sf}
\end{eqnarray}
Here $A_\mu$ ($\mu = 1,\cdots,10$) are 
traceless $N\times N$ Hermitian matrices, 
whereas $\Psi_\alpha$ ($\alpha = 1,\cdots,16$) are 
traceless $N\times N$ matrices with Grassmannian entries. 
The parameter $g$ can be scaled out by appropriate redefinition
of the matrices, and hence it is just a scale parameter.
We therefore
set $g^2 N = 1$ from now on
without loss of generality.
The integration measure for $A_\mu$ and $\Psi_\alpha$
is given by
\beq
  \dd A 
=
  \prod_{a=1}^{N^2-1} \prod_{\mu=1}^{10} \frac{\dd A_\mu^a}{\sqrt{2\pi}} \ , 
\quad \quad
  \dd \Psi
=
 \prod_{a=1}^{N^2-1} \prod_{\alpha=1}^{16} 
    \dd \Psi_\alpha^a \ ,
\label{eq:measure_psi}
\eeq
where $A_\mu^a$ and $\Psi_\alpha^a$ are 
the coefficients in the expansion 
$A_\mu = \sum_{a=1}^{N^2-1} A_\mu^a T^a$ \emph{etc.}
with respect to the $\SU(N)$ generators $T^a$ 
normalized as $\tr (T^a T^b) = \frac{1}{2}\delta^{ab}$. 

The model has an $\SO(10)$ symmetry, 
under which $A_\mu$ and $\Psi_\alpha$ transform as a vector
and a Majorana-Weyl spinor, respectively.
The $16\times 16$ matrices $\Gamma_\mu$ are the gamma matrices after 
the Weyl projection, and ${\cal C}$ is the charge conjugation matrix,
which satisfies $(\Gamma_\mu)^T = 
{\cal C} \Gamma_\mu {\cal C}^\dag$ and ${\cal C}^T = {\cal C}$.


In general, one can obtain
supersymmetric matrix models
by taking the zero-volume limit of 
pure super Yang-Mills theories
in $D=3$, 4, 6 and 10 dimensions,
where the $D=10$ case corresponds to the type IIB matrix model.
The convergence of the partition function for general $D$
was investigated
both numerically \cite{Krauth:1998xh} and analytically \cite{AW}. 
The $D=3$ model is ill-defined since the partition function
is divergent. The $D=4$ model has a real positive
fermion determinant, and Monte Carlo simulation
suggested the absence of the SSB of rotational symmetry \cite{Ambjorn:2000bf}.
(See also Refs.\ \citen{Burda:2000mn,Ambjorn:2001xs}.) 
The $D=6$ model and the $D=10$ model both
have a complex fermion determinant,
whose phase is expected to play a crucial 
role \cite{NV,Anagnostopoulos:2001yb,Anagnostopoulos:2010ux,%
Nishimura:2001sq,Nishimura:2004ts}
in the SSB of SO($D$).

In order to discuss the spontaneous symmetry breaking (SSB)
of $\SO(10)$ in the large-$N$ limit, 
we consider the 
``moment of inertia'' tensor \cite{AIKKT,Hotta:1998en}
\begin{equation}
  T_{\mu\nu} = \frac{1}{N} \tr (A_\mu A_\nu) \ , 
\label{eq:tmunu}
\end{equation}
which is a $10\times10$ real symmetric tensor. 
We denote its eigenvalues as $\lambda_j$ ($j=1,\cdots, 10$) 
with the specific order 
\begin{equation}
  \lambda_1 \geq \lambda_2 \geq \cdots \geq \lambda_{10}  \ .
\label{eq:lambda}
\end{equation}
If the SO(10) is not spontaneously broken, 
the expectation values $\langle \lambda_j  \rangle$ 
($j=1, \cdots , 10$) should be all equal in the large-$N$ limit.
Therefore, if we find that they are not equal,
it implies that the SO(10) symmetry is spontaneously broken.
Thus the expectation values $\langle \lambda_j  \rangle$ 
serve as an order parameter of the SSB.

\subsection{The Gaussian expansion method}
\label{sec:gaussian}

Since there are no quadratic terms in the actions
(\ref{eq:sb}) and (\ref{eq:sf}),
we cannot perform a perturbative expansion in the ordinary sense.
Finding the vacuum of this model is therefore a problem of 
solving a strongly coupled system.
Here we review the results obtained recently by the 
Gaussian expansion method \cite{Nishimura:2011xy}.
See also Refs.~\citen{Nishimura:2001sx,%
Kawai:2002jk}
for earlier works.
The application of such a method
to large-$N$ matrix quantum mechanics 
was advocated by Kabat and Lifschytz \cite{Kabat:2000hp},
and various black hole physics of the dual geometry 
has been discussed \cite{KLL}.


The starting point of the Gaussian expansion method
is to introduce a Gaussian term $S_0$ and to rewrite the action 
$S=S_{\rm b}+S_{\rm f}$ as 
\begin{equation}
  S = (S_0 +  S) - S_0  \ .
\label{eq:shift_s}
\end{equation}
Then we can perform a perturbative expansion
regarding the first term $(S_0 +  S)$ as
the ``classical action'' and the second term
$(-S_0)$ as the ``one-loop counter term''.
The results at finite order depend, of course, on the choice
of the Gaussian term $S_0$, which contains many free parameters
in general.
However, it is known in various examples that there exists a region
of parameters, in which the results obtained at finite order
are almost constant.
Therefore, if we can identify this ``plateau region'',
we can make concrete predictions.
It should be emphasized that the method enables us to
obtain genuinely nonperturbative results,
although most of the tasks involved are
nothing more than perturbative calculations \cite{Stevenson:1981vj}.
There are some cases in which one finds 
more than one plateau regions in the
parameter space.
In that case, each of them is considered to correspond to a local
minimum of the effective action, and
the plateau which gives the smallest free energy 
corresponds to the true vacuum. 
These statements have been confirmed explicitly
in simpler matrix models \cite{Nishimura:2002va}.

As the Gaussian action for the present model,
we consider the most general one that preserves 
the $\SU(N)$ symmetry.
Note, in particular, that 
we have to allow the Gaussian action 
to break the $\SO(10)$ symmetry
so that we can study the SSB of $\SO(10)$.
Making use of the $\SO(10)$ symmetry of the model,
we can always bring the Gaussian action into the form
\beq
\label{eq:gaussian}
  S_0 =
 \frac{N}{2} \sum_{\mu = 1}^{10} M_\mu \tr (A_\mu)^2 
+ \frac{N}{2} \sum_{\alpha,\beta=1}^{16} \mathcal{A}_{\alpha\beta} \tr 
  ( \Psi_\alpha \Psi_\beta ) \ ,
\eeq
where $M_\mu$ and $\mathcal{A}_{\alpha\beta}$ are arbitrary parameters. 
The $16\times16$ complex matrix $\mathcal{A}_{\alpha\beta}$ 
can be expanded in term of the gamma 
matrices as
%
\begin{equation}
  \mathcal{A}_{\alpha\beta} 
  = \sum_{\mu,\nu,\rho = 1}^{10} \frac{i}{3!} m_{\mu\nu\rho} 
  ({\cal C} \Gamma_{\mu} \Gamma_{\nu}^\dagger \Gamma_{\rho})_{\alpha\beta} \ ,
\label{eq:decomp_m}
\end{equation}
using a 3-form $m_{\mu\nu\rho}$.

In practice, we truncate the series expansion at some finite order.
Then the free energy 
and 
the observable 
depend on 
the free parameters $M_\mu$ and $\mathcal{A}_{\alpha\beta}$
in the Gaussian action.
%
We search for the values of parameters, at which
the free energy becomes stationary by solving 
the ``self-consistency equations''
\begin{equation}
  \frac{\partial}{\partial M_\mu} F = 0 \ , 
\qquad
  \frac{\partial}{\partial m_{\mu\nu\rho}} F = 0 \ ,
\label{eq:self-consistency}
\end{equation}
and estimate $F$ and 
observables
at the solutions.
As we increase the order of the expansion, the number
of solutions increases. If we find that there are many solutions
close to each other in the parameter space which give similar
results for the free energy and the observables, we may identify
the region as a plateau.
In fact we restrict the parameter space
by imposing the $\SO(d)$ symmetry with $2 \le d \le 7$.
The plateau region identified for each $d$ corresponds
to a local minimum which breaks the $\SO(10)$ symmetry spontaneously
down to $\SO(d)$.
By comparing the free energy,
we can determine which local minimum is actually the true vacuum.

For each $d$,
we first obtain the free energy 
up to the third order
as a function of the free parameters in the Gaussian action.
More precisely, we actually calculate ``the free energy density'' defined as
\begin{equation}
  f = 
  \lim_{N\to\infty} \left\{
  \frac{F}{N^2 - 1} 
  - \left( - 3 \log N \right) 
  \right\} \ ,
\quad
\mbox{where~}F=-\log Z \ .
\label{eq:fedensity}
\end{equation}
By differentiating the free energy density with respect to the
free parameters, we obtain
the self-consistency equations,
which we solve numerically by Mathematica.

%

\begin{figure}[tb]
\begin{center}
\begin{tabular}{cc}
\epsfig{file=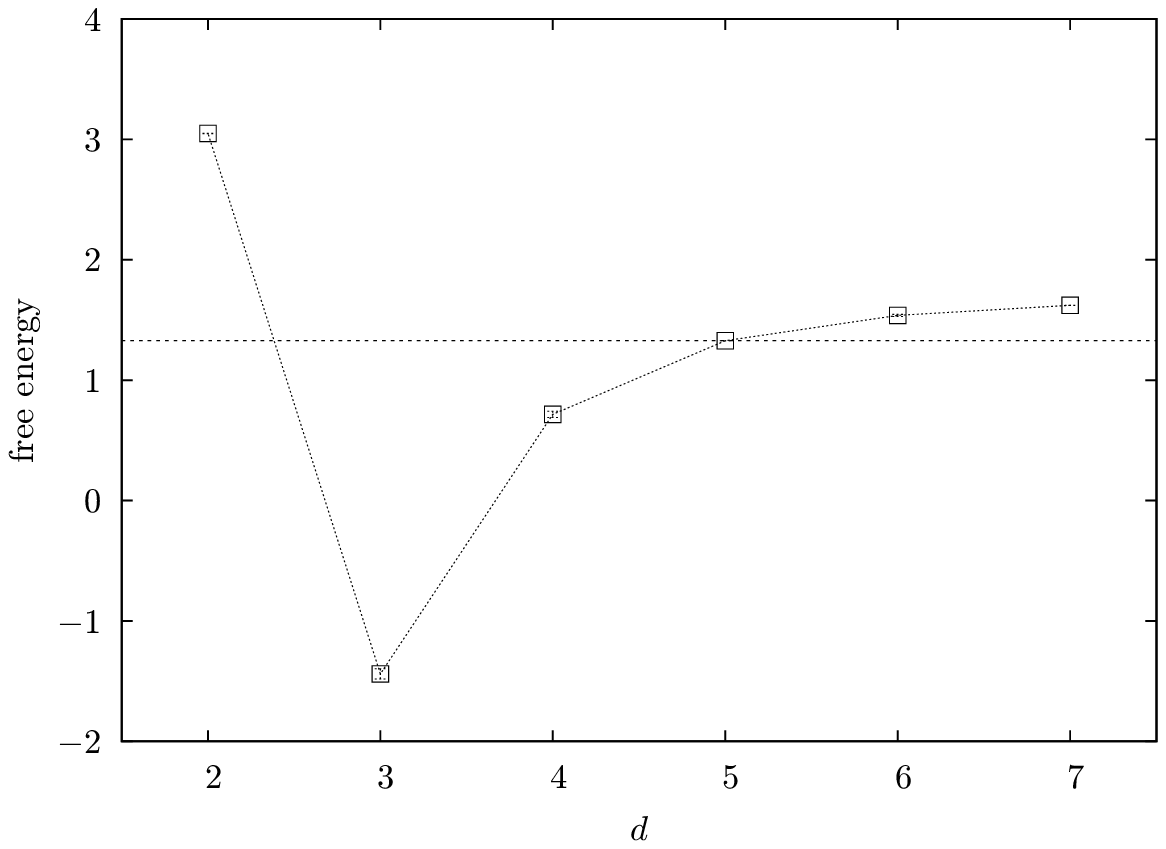,width=.47\textwidth} &
\epsfig{file=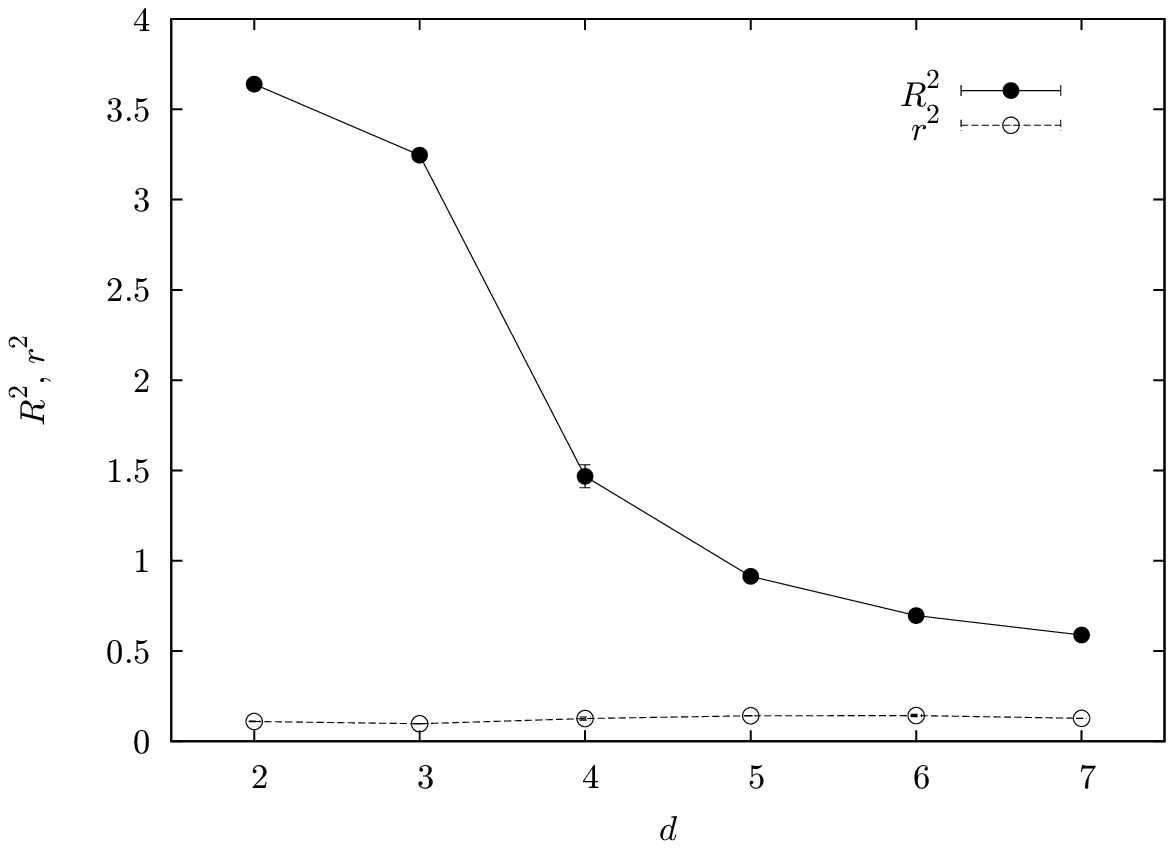,width=.47\textwidth} \\
\end{tabular}
\end{center}
\caption{
(Left) The free energy density
averaged over the ``physical solutions'' for each $d$
is plotted against $d$. 
The horizontal
line represents the KNS value $f=\log 8 - \frac{3}{4} = 1.32944\ldots$, 
and the dotted line connecting
the data points is drawn to guide the eye.
(Right) The extent of space-time $R^2$ and $r^2$ 
in the extended
and shrunken directions, respectively, 
are plotted against $d$ after taking the average
over the ``physical solutions'' for each $d$.
The solid and dashed lines connecting
the data points are drawn to guide the eye.
}
\label{fig:free-summary}
\end{figure}

In Fig.~\ref{fig:free-summary} (Left) we show
the free energy density obtained by averaging over the ``physical solutions''
(See Ref.~\citen{Nishimura:2011xy} for more detail.)
for each $d$ at order 3.
We put error bars representing the mean square error
when there are more than one physical solutions.
The horizontal line $\log 8 -\frac{3}{4}=1.32944...$
represents the result
obtained 
from the 
analytic formula of the partition function conjectured 
by Krauth, Nicolai and Staudacher (KNS) \cite{Krauth:1998xh}.

The result decreases monotonically as $d$ decreases from 7 to 3,
and it becomes much larger for $d=2$.
Thus, the $\SO(3)$ symmetric vacuum gives the smallest free energy density.
The $d$-dependence of the free energy density
is quite analogous to the one observed in the six-dimensional
case \cite{Aoyama:2010ry}.
There the value of the free energy tends to decrease
slightly as one goes from order 3 to order 5.
Considering such artifacts due to truncation,
we speculate that the KNS conjecture actually 
refers to the partition function for the ${\rm SO}(10)$ symmetric vacuum.

Let us discuss the results for the extent of space-time represented
by the eigenvalues (\ref{eq:lambda}).
For each of the $\SO(d)$ symmetric vacua, the $d$ large eigenvalues 
$\langle \lambda_j  \rangle$ $(1 \le j \le d)$
are equal due to the imposed $\SO(d)$ symmetry,
and we denote the corresponding value as $R^2$. 
The remaining $(10-d)$ eigenvalues for each
solution turn out to be quite close to each other
and we denote the mean value as $r^2$. 
In Fig.~\ref{fig:free-summary} (Right)
we plot the result for $R^2$ and $r^2$
averaged over all the physical solutions for each $d$. 
We put error bars representing the mean square error
when there are more than one physical solutions.
We find that $r^2$ stays almost constant at $r^2 = 0.1 \sim 0.15$,
which seems to be universal for all the SO($d$) symmetric
vacua with $2 \le d \le 7$.
On the other hand, the results for $R^2$
are found to
be larger for smaller $d$.
It turned out that this behavior is consistent with the 
constant-volume property \cite{Aoyama:2010ry}, which 
is given by $R^d \, r^{10-d} \approx \ell^{10}$
with $\ell^2 \approx 0.38$, except for $d=2$.




\subsection{The mechanism of SSB in the Euclidean model}

In this subsection, we review the arguments in Ref.~\citen{NV} which
show that 
the phase of the Pfaffian
favors $d(\ge 3)$-dimensional configurations
with $ \lambda_j  $ ($j=d+1, \cdots , 10$) much
smaller than the others.
This suggests the possibility that
the $\SO(10)$ symmetry is broken down to $\SO(d)$ with $d \ge 3$.


Going back to the definition of the model (\ref{eq:10dpf}),
let us first integrate over the fermionic matrices $\Psi _ \alpha$,
which yields
\beq
 \int \dd \Psi   ~ \ee ^{-S_{\rm f} } = \Pf \, {\cal M}  \ ,
\eeq
where
\beq
{\cal M}_{a \alpha ,  b \beta} = 
-  i \, f_{abc} (\,{\cal C} \, \Gamma _\mu)_{\alpha \beta} A_\mu ^c  
\eeq
is a $16 \,  (N^2-1) \times 16 \, (N^2-1)$ anti-symmetric matrix, regarding
each of $(a\alpha)$ and $(b\beta)$ as a single index.
The real totally-antisymmetric tensor $f_{abc}$ gives 
the structure constants of SU($N$).
In what follows, it proves convenient to work with an explicit 
representation of the gamma matrices given by
\beqa
&~& \Gamma _1 = i \, \sigma_2  \otimes \sigma_2  
              \otimes \sigma_2  \otimes \sigma_2 ~,~
\Gamma _2 = i \, \sigma_2  \otimes \sigma_2
              \otimes {\bf 1} \otimes \sigma_1 ~,~
\Gamma _3 = i \, \sigma_2  \otimes \sigma_2 
              \otimes {\bf 1} \otimes \sigma_3 ~,~ \n \\
&~& \Gamma _4 = i \, \sigma_2  \otimes \sigma_1
              \otimes \sigma_2 \otimes {\bf 1} ~,~
\Gamma _5 = i \, \sigma_2  \otimes \sigma_3
             \otimes \sigma_2 \otimes {\bf 1} ~,~
\Gamma _6 = i \, \sigma_2  \otimes {\bf 1}
              \otimes \sigma_1 \otimes \sigma_2  ~,~ \n \\
&~& \Gamma _7 = i \, \sigma_2  \otimes {\bf 1}
             \otimes \sigma_3 \otimes \sigma_2 ~,~
\Gamma _8 = i \, \sigma_1  \otimes {\bf 1}
              \otimes {\bf 1} \otimes {\bf 1} ~,~
\Gamma _9 = i \, \sigma_3  \otimes {\bf 1}
              \otimes {\bf 1} \otimes {\bf 1} ~,~ \n \\
&~& \Gamma _{10} = {\bf 1}  \otimes {\bf 1} \otimes {\bf 1} \otimes {\bf 1} \ ,
\label{Gamma}
\eeqa
for which the charge conjugation matrix ${\cal C}$ becomes a unit
matrix.
Note that $\Gamma_\mu$ is pure imaginary except for $\Gamma_{10}$ in this
representation. Therefore, the Pfaffian is real for $A_{10}=0$ 
(or for the Lorentzian case $A_{10} = - i A_0$).

Next, when $A_3 =A_4 = \cdots = A_{10} = 0$,
one finds that $\Pf \, {\cal M} (A)=0$ \cite{Krauth:1998xh}.
Note first that
$(\Pf \, {\cal M})^2  =\det {\cal M} = |\det U|^{16}$, where
$U$ is an $(N^2 -1)\times (N^2 -1)$ matrix defined as
$U_{ab}=f_{abc} X^c$, where $X^c = A_1^c + i \, A_2^c$.
Since $U_{ab}X^b=0$, the matrix $U$ has a zero-eigenvalue,
and therefore $\det  U = 0$.

%


Let us denote the phase of the Pfaffian by $\Gamma$.
When the configuration 
is $d(\ge 3)$-dimensional,
we find that 
\beq
\frac{\del ^n \, \Gamma}
{\del A_{\mu_1} ^{a_1} \del A_{\mu_2} ^{a_2}
\cdots \del A_{\mu_n} ^{a_n} } 
=
%
\frac{1}{2} \, 
\frac{\del ^n }{
\del A_{\mu_1} ^{a_1} \del A_{\mu_2} ^{a_2}
\cdots \del A_{\mu_n} ^{a_n}}
\,  \Imag \, \ln \det {\cal M } 
= 0  
\label{derivatives}
\eeq
for $n=1,\cdots , (9-d)$.
This is because, 
up to $(9-d)$-th order of perturbations, the configuration stays
within 9d configuration
and therefore $\det {\cal M}$ 
remains to be real positive.
This means that the phase of the Pfaffian becomes more and more 
stationary as the configuration becomes lower dimensional until
one reaches $d=3$.


\subsection{Monte Carlo studies of the Euclidean model}
\label{sec:generalized-form}

The phase of the Pfaffian makes Monte Carlo studies difficult
because of the sign problem.
However, a new method termed
the factorization method
\cite{Anagnostopoulos:2001yb,Anagnostopoulos:2010ux}
is expected to give a definite conclusion
on the SSB of SO(10) in the Euclidean model.

Let us begin by defining the distribution functions for 
the eigenvalues $\{ \lambda_k \}$
as 
\beqa
\label{rho_full_gen1}
\rho(x_1 , \cdots  , x_{10} ) &=& 
\Big\langle \prod_{k}\delta (x_k -  \lambda_{k}) \Big\rangle \ , \\
\rho^{(0)}(x_1 , \cdots  , x_{10} ) &=& 
\Big\langle \prod_{k}\delta (x_k - \lambda_{k}) \Big\rangle_0 
\label{rho_full_gen}
\eeqa
for the full model and the phase-quenched model, respectively.
By definition (\ref{eq:lambda}),
these functions vanish unless
$x_1 \ge \cdots  \ge x_{10}$.
Applying the reweighting formula to the right-hand side of
(\ref{rho_full_gen1}),
one finds that it factorizes as
\beq
\label{eq:rhodef_gen}
\rho(x_1 , \cdots  , x_{10})
 = \frac{1}{C} \, 
\rho^{(0)}(x_1 , \cdots  , x_{10})
 \, w(x_1 , \cdots  , x_{10}) \ .
\eeq
The real parameter $C$ is 
a normalization constant given by\footnote{In the second equality,
we have used the fact that the phase $\Gamma$ flips its sign
under the parity transformation $A_{10} \mapsto - A_{10}$,
which is a symmetry of the phase-quenched model.
} 
\begin{equation}
\label{eq:cconst}  
C \stackrel{\textrm{def}}{=} 
  \langle e^{i \Gamma} \rangle_{0} = 
  \langle \cos \Gamma \rangle_{0} \ ,
\end{equation}
which need \emph{not} be calculated in the present method.
The function $w(x_1 , \cdots  , x_{10})$ is defined by
\begin{equation}
w(x_1 , \cdots  , x_{10}) 
\stackrel{\textrm{def}}{=} 
  \langle e^{i \Gamma} \rangle_{x_1 , \cdots  , x_{10}} = 
  \langle \cos \Gamma  \rangle_{x_1 , \cdots  , x_{10}} \ , 
\label{w_n_gen}
\end{equation}
where $\langle \  \cdot  \ \rangle_{x_1 , \cdots  , x_{10}}$ 
denotes a VEV with respect to
the partition function
\begin{equation}
 Z_{x_1 , \cdots  , x_{10}} 
= \int dA \, \ee^{-S_{0}} \, 
\prod_{k} \delta (x_k -\lambda_{k})
\ .
\label{ix_pf_gen}
\end{equation}

At large $N$, the expectation values $\langle \lambda _ k \rangle$
are obtained by 
maximizing the
distribution function $\rho(x_1 , \cdots  , x_{10})$
with respect to $x_1 , \cdots , x_{10}$.
This leads to the coupled equations
\beq
\frac{\del }{\del x_n} 
\log \rho^{(0)}(x_1 , \cdots  , x_{10})  = 
- \frac{\del }{\del x_n}  \log w(x_1 , \cdots  , x_{10}) 
  \quad \mbox{for $n=1, \cdots , 10$}  \ .
\label{general-master-eq}
\eeq
The function on the left-hand side and 
$w(x_1 , \cdots  , x_{10})$ defined by (\ref{w_n_gen})
can be obtained by simulating the constrained model (\ref{ix_pf_gen}).

An important observation now \cite{Anagnostopoulos:2001yb} is that 
$\log w(x_1 , \cdots  , x_{10}) \sim {\rm O}(N^2)$ as naturally
expected from the number of degrees of freedom.
On the other hand, we speculate \cite{Aoyama:2010ry} that
the left-hand side (\ref{general-master-eq}) is O($N$)
if $\prod x_k = (\ell^2)^{10}$ and $x_{n} \ge r^2$ , where 
$\ell$ and $r$ are some scale determined by the phase-quenched model.
Thus the maximum is essentially given by
$x_1 = \cdots = x_d=R^2$ and $x_{d+1} = \cdots = x_{10}= r^2$
with $R^d \, r^{10-d} = \ell^{10}$, where $3 \le d \le 9$.
In order to determine which $d$ gives the absolute maximum of 
$\rho(x_1 , \cdots  , x_{10})$, 
one can neglect the effect of $\rho^{(0)}(x_1 , \cdots  , x_{10})$ 
since it is suppressed by $1/N$.
Due to the property in the previous section,
it is expected that $\log w(x_1 , \cdots  , x_{10})$ becomes maximum 
for the solution with $d=3$.
Therefore, $\rho(x_1 , \cdots  , x_{10})$ becomes maximum also for
$d=3$. It is anticipated that
the results of the Gaussian expansion method
can be reproduced by Monte Carlo simulation in this way.

\subsection{Definition of the Lorentzian model}

In this section we discuss the Lorentzian matrix model
as a nonperturbative formulation of
type IIB superstring theory in (9+1)-dimensional space-time \cite{Kim:2011cr}.
Our starting point is the action \cite{IKKT} $S = S_{\rm b} + S_{\rm f}$, where 
\beqa
S_{\rm b} &=& -\frac{1}{4g^2} \, \tr \Bigl( [A_{\mu},A_{\nu}]
[A^{\mu},A^{\nu}] \Bigr) \ , \n  \\
S_{\rm f}  &=& - \frac{1}{2g^2} \,
\tr \Bigl( \Psi _\alpha (\, {\cal C} \,  \Gamma^{\mu})_{\alpha\beta}
[A_{\mu},\Psi _\beta] \Bigr)  \ ,
\label{action}
\eeqa
with  $A_\mu$ ($\mu = 0,\cdots, 9$) and
$\Psi_\alpha$ ($\alpha = 1,\cdots , 16$) being
$N \times N$ traceless Hermitian matrices.
The Lorentz indices $\mu$ and $\nu$ are 
raised and lowered
using the metric
$\eta={\rm diag}(-1 , 1 , \cdots , 1)$.
The $16 \times 16$ matrices $\Gamma ^\mu$ are
ten-dimensional gamma matrices after the Weyl projection,
and the unitary matrix ${\cal C}$ is the
charge conjugation matrix.
The action has manifest SO(9,1) symmetry,
where $A_{\mu}$ and $\Psi _\alpha$ transform as a
vector and
a Majorana-Weyl spinor, respectively.
The Euclidean model,
which has SO(10) symmetry,
can be obtained from this action
by the Wick rotation $A_0=i A_{10}$, $\Gamma^0 = - i \Gamma_{10}$.
A crucial difference is that
the bosonic part of the action in the Euclidean model
is positive definite, whereas in the Lorentzian model it is
\beq
\tr (F_{\mu\nu}F^{\mu\nu})=
- 2 \, \tr (F_{0i})^2 + \tr (F_{ij})^2 \ ,
\label{bosonic-action}
\eeq
where $F_{\mu\nu} = - i [A_\mu , A_\nu]$ are Hermitian matrices,
and hence the two terms in (\ref{bosonic-action}) have opposite signs.


We define the partition function of the Lorentzian model by
\beq
Z = \int d A \, d\Psi \, e^{i S} =
\int d A \,  e^{i S_{\rm b}} {\rm Pf}{\cal M}(A)
\ ,
\label{partition-fn-def}
\eeq
where the Pfaffian ${\rm Pf}{\cal M}(A)$ appears from integrating
out the fermionic matrices $\Psi_\alpha$.
Note that in the Euclidean model,
the Pfaffian is complex in general,
and its phase plays a crucial role in the 
SSB
of SO(10) symmetry \cite{NV,Anagnostopoulos:2001yb,Anagnostopoulos:2010ux,%
Nishimura:2001sq,Nishimura:2004ts}.
On the other hand, the Pfaffian in the Lorentzian model is \emph{real}.
Therefore, the mechanism of SSB that was identified
in the Euclidean model is absent in the Lorentzian model.

In the definition (\ref{partition-fn-def}),
we have replaced the ``Boltzmann weight'' $e^{-S}$
used in the Euclidean model by $e^{iS}$.
This is theoretically motivated
from the connection
to the worldsheet theory \cite{IKKT}.
The partition function
(\ref{partition-fn-def})
can also be obtained formally
from pure ${\cal N}=1$ supersymmetric Yang-Mills theory
in $(9+1)$ dimensions by dimensional reduction.
Note, however, that
the expression (\ref{partition-fn-def}) is ill-defined and
requires appropriate regularization
in order
to make any sense out of it.
It turns out that the integration over $A_\mu$ is divergent,
and we need to introduce two constraints
\beqa
\frac{1}{N}\tr (A_0)^2  &\le&  \kappa \frac{1}{N} \tr (A_i)^2  \ ,
\label{T-constr} \\
\frac{1}{N} \tr (A_i)^2   &\le&  L^2  \ .
\label{R-constr}
\eeqa
This is in striking contrast to the Euclidean model,
in which the partition function is shown to
be finite without any regularization \cite{Krauth:1998xh,AW}.

Note that $e^{iS_{\rm b}}$ in the partition function (\ref{partition-fn-def})
is a phase factor
just as in the path-integral formulation of quantum field theories
in Minkowski space.
However, we can circumvent the sign problem by integrating out the scale factor
of $A_\mu$, which essentially replaces the phase $e^{iS_{\rm b}}$ by
the constraint $S_{\rm b} \approx 0$. (Such a constraint is analogous
to the one that appeared in the model inspired by space-time uncertainty
principle \cite{Yoneya:1997gs}.)
Without loss of generality, we set $L=1$ in (\ref{R-constr}), 
and thus we arrive at the model
\beq
Z =  \int dA \,
\delta \left(
\frac{1}{N}\tr (F_{\mu\nu}F^{\mu\nu})  \right)
 {\rm Pf} {\cal M} (A)
\, \delta\left(\frac{1}{N}\tr (A_i)^2 - 1 \right)
\theta\left(\kappa  - \frac{1}{N}\tr (A_0)^2  \right)  \ ,
\label{our-model}
\eeq
where $\theta(x)$ is the Heaviside step function.
Since the Pfaffian ${\rm Pf} {\cal M}(A)$ is real
in the present Lorentzian case,
the model (\ref{our-model}) can be studied by Monte Carlo simulation
without the sign problem.\footnote{Strictly speaking, the Pfaffian
can flip its sign, but we find that the configurations with
positive Pfaffian dominates as $N$ is increased.
Hence, we just take the absolute value of the Pfaffian in actual simulation.} 
Note that this is usually not the case for quantum field theories in
Minkowski space.
\subsection{Monte Carlo studies of the Lorentzian model}
\label{sec:MClorentz}

We perform Monte Carlo simulation of the model (\ref{our-model})
by using the 
RHMC
algorithm \cite{Clark:2003na} as in the 1d SYM case discussed 
in section \ref{sec:MCgg}.

\begin{figure}[tb]
\begin{center}
\begin{tabular}{cc}
\epsfig{file=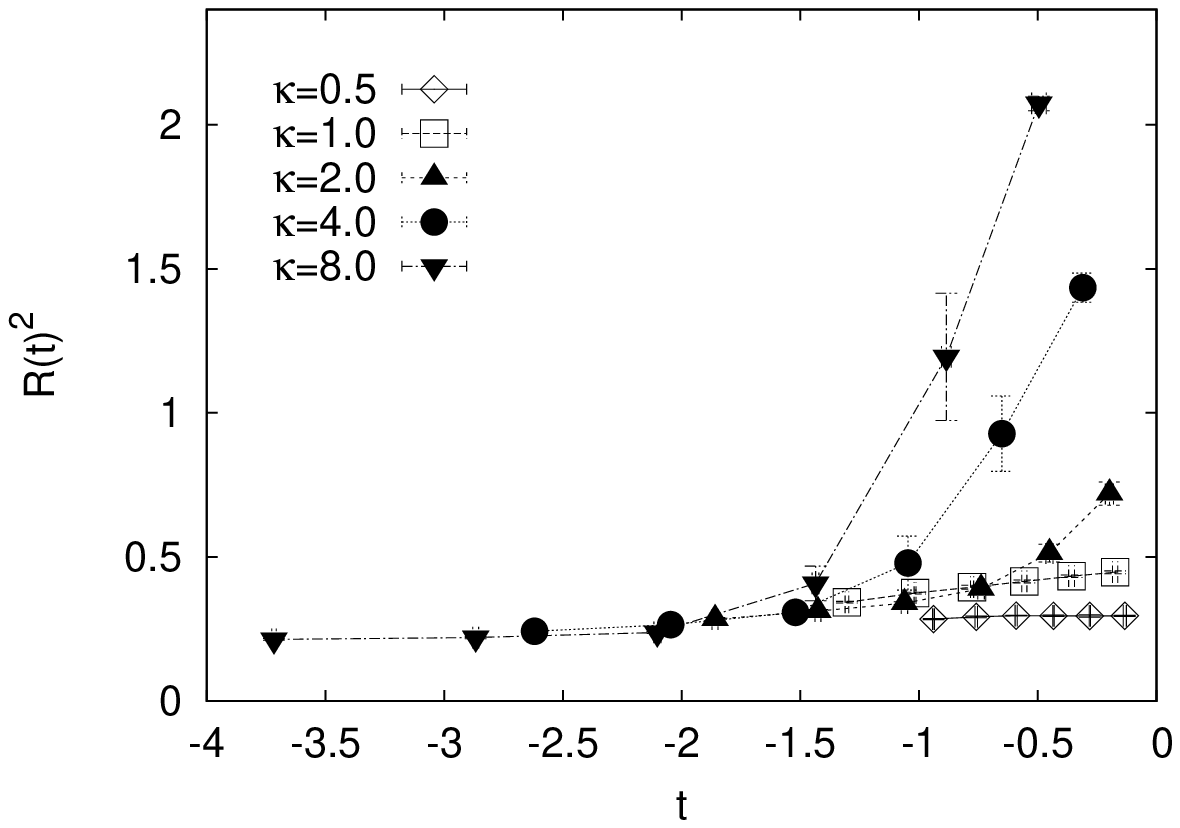,width=.47\textwidth} &
\epsfig{file=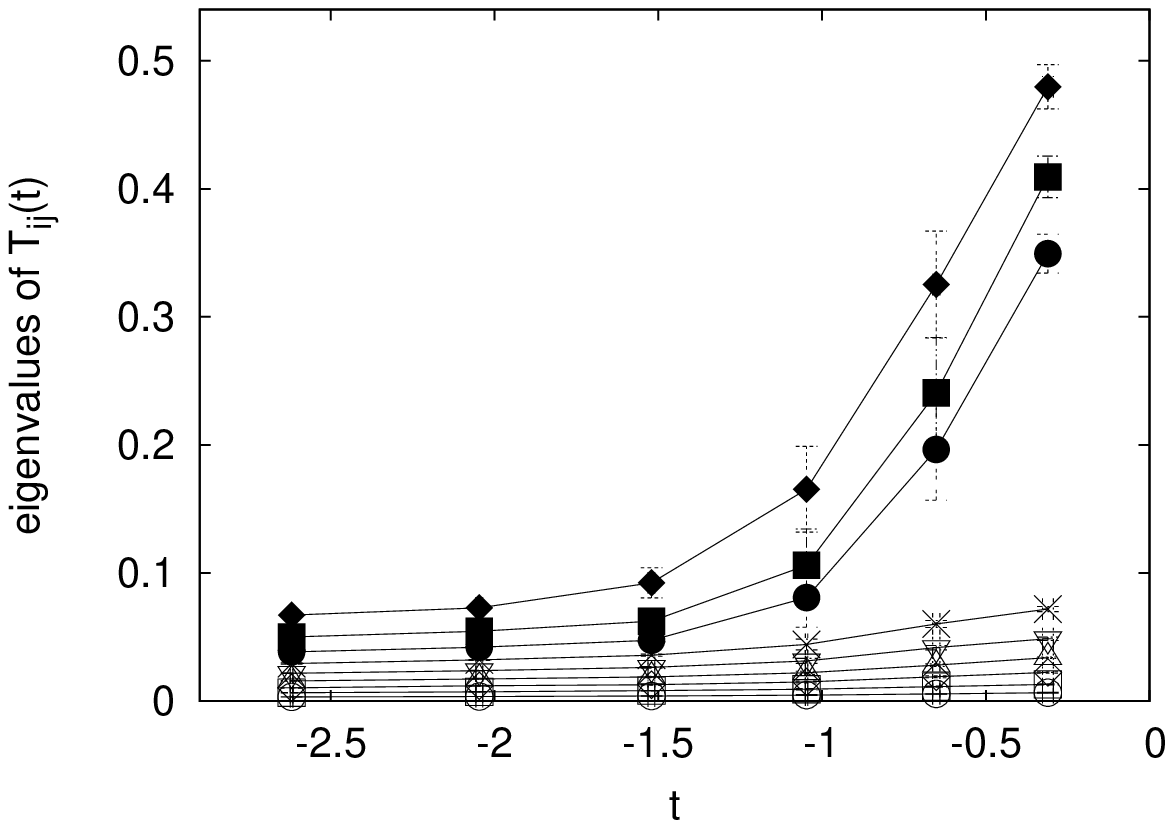,width=.47\textwidth} \\
\end{tabular}
\end{center}
\caption{
(Left) The extent of space $R(t)^2$ with $N=16$ and $n=4$
is plotted as a function of $t$
for five values of $\kappa$.
The peak at $t=0$ starts to grow at some critical $\kappa$.
(Right) The 9 eigenvalues of $T_{ij}(t)$ with $N=16$ and $n=4$
are plotted as a function of $t$
for $\kappa=4.0$. After the critical time $t_{\rm c}$,
3 eigenvalues become larger,
suggesting that the SO(9) symmetry is spontaneously broken
down to SO(3).
}
\label{Rt}
\end{figure}

In order to extract the ``time evolution'', we diagonalize $A_0$,
and define the eigenvectors $| t_a \rangle$ corresponding
to the eigenvalues $t_a$ of $A_0$ ($a=1 , \cdots , N$)
with the specific order $t_1 < \cdots < t_N$.
The spatial matrix in this basis $\langle t_{a} | A_i | t_{b} \rangle $
is not diagonal, but it turns out that the off-diagonal elements
decrease rapidly as one goes away from a diagonal element.
This motivates us to
define $n\times n$ matrices
$\bar{A}_i^{(ab)}(t) \equiv  \langle t_{\nu+a} | A_i | t_{\nu+b} \rangle $
with $1 \le a , b \le n$ and
$t= \frac{1}{n}\sum_{a=1}^{n} t_{\nu + a}$
for $\nu=0,\cdots , (N-n)$.
These matrices represent the 9d space structure 
at fixed time $t$.
(This point of view can be justified in the large-$N$ limit,
in which more and more eigenvalues of $A_0$ 
appear around some value $t$
within a fixed interval $\delta t$.)
The block size $n$ should be large enough to include non-negligible 
off-diagonal elements.
In Fig.~\ref{Rt} (Left) we plot the extent of space 
$R(t)^2 \equiv \frac{1}{n} \tr
\bar{A}_i(t)
^2$ for $N=16$ and $n=4$.
Since the result is symmetric under the time reflection
$t \rightarrow -t$ as a consequence of the symmetry $A_0 \rightarrow -A_0$,
we only show the results for $t<0$.
There is a critical $\kappa$, beyond which
the peak at $t=0$ starts to grow.

Next we study the spontaneous breaking of the SO(9) symmetry.
As an order parameter,
we define the $9 \times 9$
(positive definite)
real symmetric tensor
\beq
T_{ij}(t) = \frac{1}{n} \tr \Bigl\{
\bar{A}_i(t) \bar{A}_j(t) \Bigr\} \ ,
\eeq
which is an analog of (\ref{eq:tmunu})
in the Euclidean model.
The 9 eigenvalues of $T_{ij}(t)$ are
plotted against $t$ in Fig.~\ref{Rt} (Right) for $\kappa=4.0$.
We find that 3 largest eigenvalues of
$T_{ij}(t)$ start to grow at the critical
time $t_{\rm c}$, which suggests that the SO(9)
symmetry is spontaneously broken down to
SO(3) after $t_{\rm c}$.
Note that $R(t)^2$ is given by the sum of 9 eigenvalues
of $T_{ij}(t)$.

\begin{figure}[tb]
\begin{center}
\begin{tabular}{cc}
\epsfig{file=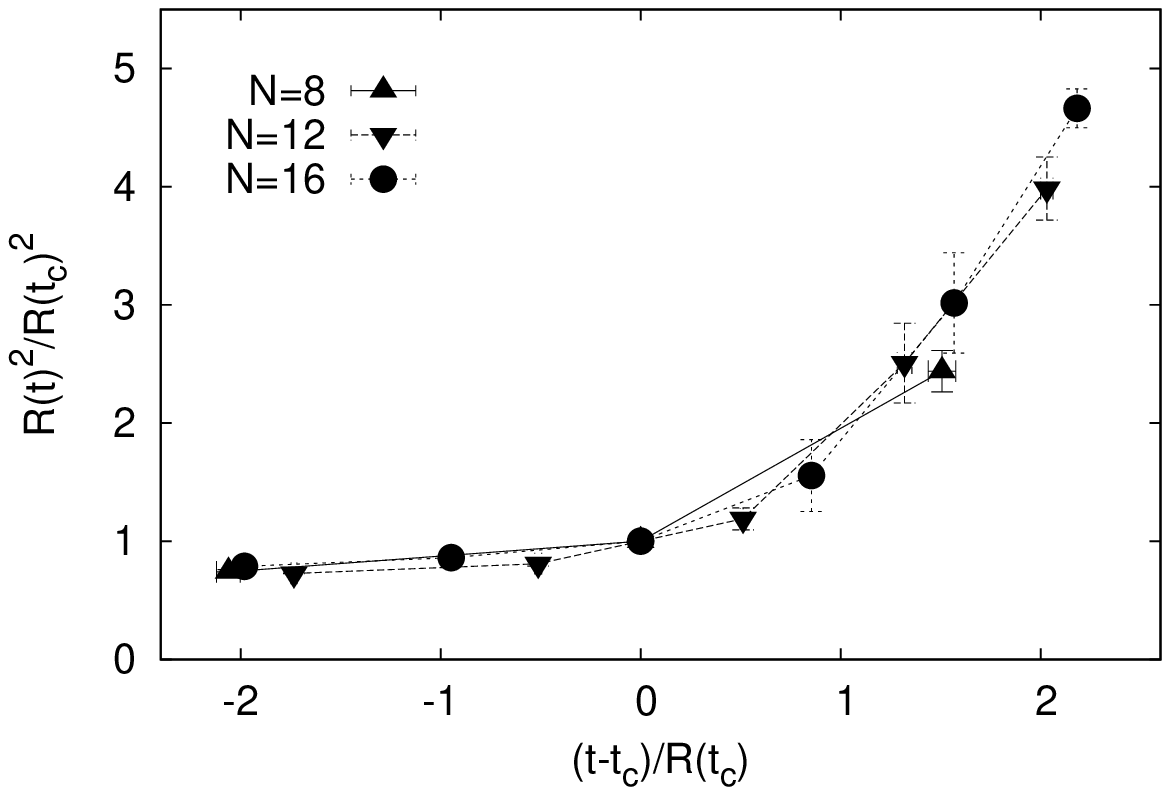,width=.47\textwidth} &
\epsfig{file=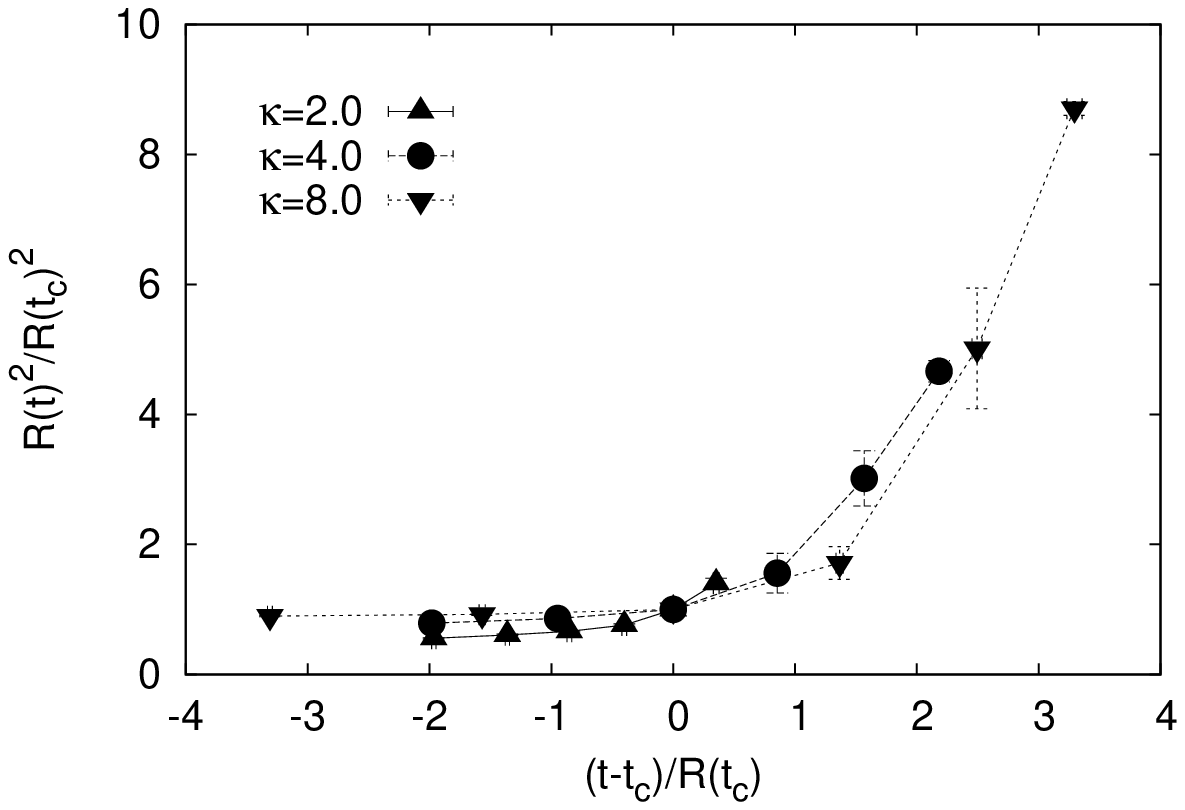,width=.47\textwidth} \\
\end{tabular}
\end{center}
\caption{
(Left) The extent of space $R(t)^2$ 
for $\kappa=\beta N^{1/4}$ is plotted for $N=8,12,16$ with $\beta=2$.
We plot the results against the
shifted time $t - t_{\rm c}$ in units of
the size of the universe $R(t_{\rm c})$
at the critical time.
(Right) Similar plot for fixed $N=16$ with $\kappa=2,4,8$.
}
\label{Rt-rescaled}
\end{figure}

It turned out that one can remove the
infrared cutoffs $\kappa$ and $L$ 
in the large-$N$ limit
in such a way that $R(t)$ scales.
This can be done in two steps.
(i) First we send $\kappa$ to $\infty$ with $N$ as
$\kappa = \beta \, N^{p}$ ($p\simeq \frac{1}{4}$) \cite{KNT}.
The scaling behavior is clearly seen in Fig.~\ref{Rt-rescaled} (Left).
The scaling curve of $R(t)$ one obtains in this way
depends on $\beta$.
(ii) Next we send $\beta$ to $\infty$ with $L$.
The two limits correspond to
the continuum limit and
the infinite volume limit, respectively, in quantum field theory.
Thus the two constraints (\ref{T-constr}), (\ref{R-constr})
can be removed in the large-$N$ limit,
and 
the resulting theory has no parameter other than one scale parameter.

Let us discuss the second limit (ii) in more detail.
We find that the inequality (\ref{R-constr}) is actually saturated 
for the dominant configurations.
Therefore, one only has to make the rescaling $A_\mu \mapsto L A_\mu$
in order to translate the configurations in the model (\ref{our-model})
as those in the original partition function.
It turns out that $R(t)$ for the rescaled configurations scales
in $\beta$ by tuning $L$ and shifting $t$ appropriately.
In order to see this, it is convenient to 
choose $L$ so that $R(t)$ 
at the critical time $t=t_{\rm c}$ becomes unity,
and to shift $t$ so that the critical time comes to the origin.
Then $R(t)$ with increasing $\beta$ 
extends in $t$
in such a way that the results at smaller $|t|$ scale.
This is demonstrated in Fig.~\ref{Rt-rescaled} (Right),
where we find a reasonable scaling behavior
for $N=16$ with $\kappa=2.0, 4.0, 8.0$.
Note, in particular, that the extent of ``time'' increases
as $\kappa$ is increased, which is not the case in the bosonic model \cite{KNT}.
Thus, supersymmetry is crucial for the ``emergent time'' 
in the Lorentzian matrix model.
\subsection{The mechanism of SSB in the Lorentzian model}

The SSB of SO(9) looks mysterious at first sight, but
we can actually understand the mechanism quite intuitively.
Let us consider the case in which $\kappa$ is large.
Then the first term of (\ref{bosonic-action}) becomes
a large negative value, and therefore the second term
has to become large in order to 
make (\ref{bosonic-action}) zero as required in (\ref{our-model}).
Due to the constraint $\frac{1}{N} \tr (A_i)^2=1$, however,
it is more efficient to maximize the second term of (\ref{bosonic-action})
at some fixed time.
The system actually chooses the middle point $t=0$, where
the suppression on $A_i$ from the first term of (\ref{bosonic-action})
becomes the least.
This explains why the peak of $R(t)$ at $t=0$ grows as we increase $\kappa$.

Let us then consider a simplified question: what is the configuration of $A_i$
which gives the maximum $\frac{1}{N} \tr (F_{ij})^2$ with fixed
$\frac{1}{N} \tr (A_i)^2=1$. Using the Lagrange multiplier $\lambda$,
we maximize the function $G=  \tr (F_{ij})^2 - \lambda \,  \tr (A_i)^2$.
Taking the derivative with respect to $A_i$, we obtain
$2\, [A_j,[A_j, A_i]] - \lambda A_i = 0 $.
This equation can be solved if $A_i = \chi L_i$ for $i\le d$,
and $A_i = 0$ for $d < i \le 9$,
where $L_i$ are the representation matrices
of a compact semi-simple Lie algebra with $d$ generators.
Clearly $d$ should be less than or equal to 9.
It turns out that the maximum of
$\frac{1}{N} \tr (F_{ij})^2$ is achieved for the SU(2) algebra,
which has $d=3$,
with $L_i$ being the direct sum of the spin-$\frac{1}{2}$ representation
and $(N-2)$ copies of the trivial representation.
This implies the SSB of SO(9) down to SO(3).
The SSB
can thus be understood
as a classical effect in the $\kappa\rightarrow \infty$ limit.
%
When we tune $\kappa$ with increasing $N$ as described above,
quantum effects become important.
We have confirmed \cite{KNT} that the $n\times n$ matrix
$Q = \sum_{i=1}^9 \bar{A}_i(t)^2$ has 
quite a continuous eigenvalue distribution,
which implies that the space is not like a two-dimensional sphere as
one might
suspect from the classical picture.




\section{Summary}
\label{sec:sum-all}

We have discussed the origin of space-time
from the viewpoint that matrices are the fundamental degrees of
freedom in superstring theory.
We have seen that
matrices indeed
provide an appropriate description of the quantum space-time
at the singularities that appear inside a
black hole or at the beginning of the Universe.

In particular,
Monte Carlo studies of U($N$) SYM
have deepened our understanding
on the gauge/gravity duality considerably.
In the D0-brane case, it would be interesting to investigate
whether the duality holds including the $1/N$ corrections,
which should be compared with the string loop corrections on the
gravity side.
The D3-brane case, which corresponds to the 4d ${\cal N}=4$ SYM,
is challenging, but we hope that more nontrivial tests of 
the AdS/CFT correspondence are possible by measuring non-circular 
Wilson loops \cite{Honda:2011qk}
and four-point functions \cite{Arutyunov:2000py}.
As a new direction, Monte Carlo simulation of matrix models 
obtained by applying the localization technique \cite{Pestun:2007rz}
to supersymmetric gauge theories 
is expected to be useful in extending the first-principle studies of
the gauge/gravity duality to many more examples \cite{Hanada:2012si}.

The results for the Lorentzian matrix model, on the other hand,
suggest that (3+1)-dimensional expanding
universe emerges dynamically from type IIB superstring
theory if the theory is treated nonperturbatively.
This may be contrasted with the quantum cosmology in the early 80s that
aimed at describing the birth of the universe \cite{vilenkin}
within the mini-superspace approximation.\footnote{More recently, 
a nonperturbative approach to quantum gravity
has been pursued
using the causal dynamical triangulation \cite{Ambjorn:2005qt}.
For earlier works that put forward the idea to use matrices
for cosmology, see Refs.~\citen{Freedman:2004xg}.
See also Refs.~\citen{ncym} for related works on emergent gravity.}
Note also that 
the picture suggested here
is quite different from that 
in (perturbative) superstring theory, where space-time
with various dimensions can be obtained 
by compactification or by using D-brane backgrounds.

The rapid expansion of the three-dimensional space
observed in Monte Carlo simulation 
may be 
interpreted
as the beginning of inflation.
It would be interesting to investigate
the microscopic origin of the inflation 
along this line. 
Since the mechanism of the SSB relies crucially on the 
noncommutativity of space, it is important to see
how a commutative space-time appears at later times.
Furthermore, it might be possible to understand the origin of
dark energy found in the present cosmological observations
and to predict the fate of our Universe by studying the
late time physics of the Lorentzian model.

%
Since superstring theory is not only a theory of quantum gravity
but also a theory of all the matters and the fundamental 
interactions among them,
it would be interesting to see how the Standard Model
appears at later times in
the Lorentzian matrix model.
Finding solutions to the classical equation of motions
\cite{Aoki:2010gv,Steinacker:2011wb,Chatzistavrakidis:2011gs,%
Chatzistavrakidis:2011su}
and performing perturbative expansion around them
would be an important direction as an approach complementary to
Monte Carlo simulation.


We hope that all the ideas and technologies that
we learned 
in Monte Carlo studies of QCD would
be useful in developing the Monte Carlo studies of 
superstring theory further.






\section*{Acknowledgements}
We would like to thank M.\ Honda
and A.\ Tsuchiya for 
useful comments on the manuscript.
This work is supported 
by Grant-in-Aid for Scientific
Research (No.\ 20540286 and 23244057) 
from Japan Society for the Promotion of Science.

%

\end{document}